\documentclass[a4paper,11pt]{article}

\usepackage{amssymb,amsmath,amsfonts}
\usepackage{float}
\usepackage{graphicx}
\usepackage{tikz}
\usetikzlibrary{shapes, arrows, shadows}
\usepackage{upgreek}

\def\be{\begin{equation}}
\def\ee{\end{equation}}
\def\bea{\begin{eqnarray}}
\def\eea{\end{eqnarray}}

\begin{document}

\begin{titlepage}
\date{\today}       \hfill

\begin{center}

\vskip .5in

{\LARGE \bf   Critical Ising Model with Boundary Magnetic Field: RG Interface and Effective Hamiltonians }\\
\vspace{5mm}

\today
 
\vskip .250in

\vskip .5in
{\large Anatoly Konechny}

\vskip 0.5cm
{\it Department of Mathematics,  Heriot-Watt University\\
Edinburgh, EH14 4AS, United Kingdom\\[10pt]
and \\[10pt]
Maxwell Institute for Mathematical Sciences\\
Edinburgh, EH14 4AS, United Kingdom\\[10pt]
}
E-mail: A.Konechny@hw.ac.uk
\end{center}

\vskip .5in
\begin{abstract} \large
Critical 2D Ising model with a boundary magnetic field is arguably the simplest QFT that 
interpolates between two non-trivial fixed points. We use the diagonalising Bogolyubov transformation 
for this model to investigate two quantities. Firstly we explicitly construct  an RG interface operator that is  a boundary condition changing operator 
linking the free boundary condition with the one with a boundary magnetic field. We investigate its properties and in particular 
show that in the limit of large magnetic field this 
operator becomes the dimension 1/16  primary field linking the free and fixed boundary conditions. Secondly we use Schrieffer-Wolff method 
to construct  effective Hamiltonians both near the UV and IR fixed points. 
\end{abstract}

\end{titlepage}

\renewcommand{\thepage}{\arabic{page}}
\setcounter{page}{1}
\large 

\section{Introduction }
\renewcommand{\theequation}{\arabic{section}.\arabic{equation}}
We consider the two-dimensional critical Ising model on an upper  half-plane with complex coordinates $z=x+iy$, $\bar z=x-iy$ and 
the boundary at $z=\bar z$.
The model admits two elementary conformal boundary conditions corresponding to keeping 
the boundary spin fixed (up or down) or allowing it to fluctuate freely. In terms of free massless fermions 
$\psi(z)$, $\bar \psi(\bar z)$ the free boundary condition corresponds to setting $\psi(z)=\bar \psi(\bar z)$, $z=\bar z$ 
and the fixed one corresponds to $\psi(z)=-\bar \psi(\bar z)$, $z=\bar z$. 
For the free boundary condition the vacuum is doubly degenerate. In the quantisation with time along the boundary we can choose 
a basis of spin up and spin down vacua: $|0,\pm\rangle$.
To describe  the boundary spin operator 
we introduce following \cite{GZ}, \cite{CZ} a boundary fermion field $a(x)$  with a two-point function 
\be
\langle a(x) a(x')\rangle = \frac{1}{2}{\rm sign}(x-x') \, .
\ee
The corresponding operator $a$ acts on the vacua as $a|0,\pm\rangle = |0,\mp\rangle$. 

Starting with a free boundary condition we can perturb it by switching on 
  a boundary magnetic field $h$ that couples to the boundary spin operator $\sigma_{B}(x)=i(\psi(x) + \bar \psi(x))a(x)$.  
The Euclidean action functional describing the perturbed model on the upper half plane is 
\be
S=\frac{1}{2\pi}\int\limits_{-\infty}^{\infty}\!\!dx\int\limits_{0}^{\infty}\!\!dy[\psi  \partial_{\bar z}\psi + \bar \psi \partial_{z}\bar \psi] 
+ \int\limits_{-\infty}^{\infty}\!\!dx \Bigl[ -\frac{i}{4\pi}\psi \bar \psi + \frac{1}{2}a\partial_{x}a + ih(\psi + \bar \psi)a \Bigr] \, .
\ee

This model was introduced   in \cite{GZ} where in particular  it was 
shown that it is integrable. The local magnetisation was computed in \cite{CZ} and the exact boundary state was found 
in \cite{Chat}. (The results of two latter papers were also generalised to non-critical temperature in \cite{Chat2}, \cite{Mir}
 but in the present paper we confine ourselves to the critical bulk.) The boundary entropy flow was studied in \cite{me1} while   
 the approach to the infrared fixed point and some numerical aspects were studied in \cite{Toth}, \cite{Toth2}. In the latter two papers 
 it was shown how to solve the model in the Hamiltonian formalism using a Bogolyubov transformation.
 
 In the present paper our main interest in this model lies in the fact that while being Gaussian and solvable by fairly elementary methods 
 it interpolates between two non-trivial fixed points. This allows one to use it as a laboratory for investigating the approach 
 to non-trivial fixed points. While the onset of RG flow near a UV fixed point has been studied extensively in QFT literature, and has many 
 results which are now considered standard, the 
situation near a nontrivial  end point of the flow is not so well understood.  We focus our attention on two quantities. The first is the RG interface. 
 The idea of an RG interface was first spelled out in \cite{FQ}, \cite{BR} and then elaborated in \cite{BR} and \cite{Gaiotto}. 
In general, for a perturbation of a Euclidean UV CFT by a relevant operator $\phi(x)$ with coupling $\lambda$ 
we consider a set up in which we perturb 
 on a half space and renormalise it by adding the usual counter terms inside the perturbed region as well as, possibly, 
 additional counter terms on the boundary of the perturbed region. 
 The renormalised theory defines a codimension one extended object which we call an interface between 
the UV CFT  and the perturbed QFT. The interface has its own stress-energy tensor conserved together with the stress-energy tensors of 
the bulk theories \cite{Billo_etall}, \cite{BulkBoundary}. If we take the RG scale on the perturbed side to infinity we get a conformal interface between 
the UV CFT  and the IR CFT. The latter can be trivial in which case we get a conformal boundary condition describing the vacuum of the perturbed theory 
in the far infrared. 

For a boundary CFT in two-dimensions (BCFT) perturbed by a boundary operator the corresponding interface between the UV BCFT  
and the perturbed theory is point-like and is thus described by a local boundary condition changing operator.  We 
denote this operator $\hat \psi_{\lambda, 0} $ assuming that we are on the upper half plane and  the perturbed boundary condition is to the right of $\hat \psi_{\lambda, 0} $ (see the picture below). 
Similarly we introduce a  conjugate operator $\hat \psi_{0, \lambda}=\hat \psi_{\lambda, 0}^{\dagger}$  that has the perturbed boundary condition 
on its right.

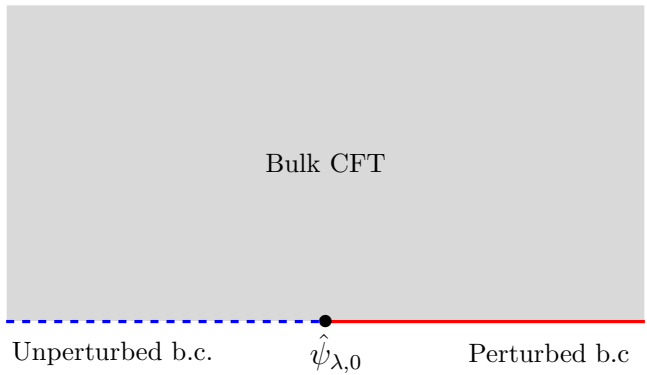
\begin{center}
\begin{tikzpicture}[scale=1.4]
\filldraw[fill=gray!30!white,draw=white] (-3,0) rectangle (3,3);
\draw[blue,very thick,dashed](-3,0)--(0,0);
\draw[red,very thick] (0,0)--(3,0);
\draw (0,0) node {$\bullet$} ;
\draw (0.1,-0.3) node {$\hat \psi_{\lambda, 0}$};
\draw (0,1.5) node {{\small Bulk CFT}};
\draw (-2,-0.3) node {{\small Unperturbed b.c.}};
\draw (2.1, -0.3) node {{\small Perturbed b.c}} ;
\end{tikzpicture}
\end{center}

Any boundary RG flow 
always ends at some infrared BCFT (there is no mass gap and no boundary analogue of the trivial QFT). We thus get a boundary condition 
changing operator between two BCFTs which we call an RG  operator.  
Unlike in the case of bulk RG flows where (with the exception of very few cases) 
we don't have a description of all conformal interfaces between the UV and IR CFTs, for two BCFT's (with the same two-dimensional bulk CFT) 
we often have a complete description of all boundary condition changing operators. For example this is the case when the bulk CFT is a 
Virasoro minimal model. Then all boundary conditions and all boundary condition changing operators have been classified by Cardy 
\cite{Cardy}. Cardy's construction has been generalised in \cite{BPPZ} to the case of general rational CFTs. For boundary RG flows 
in the minimal models triggered by $\psi_{1,3}$ boundary operators a concrete proposal for an RG operator between the UV and IR BCFTs 
was investigated in \cite{me2}.

In the present paper we explicitly construct the RG interface operator for the boundary magnetic flow. It is given by formula (\ref{psi_hat1}) which 
is one of our main results. For the infinite magnetic field limit we show that this operator tends to the boundary condition changing primary 
field of scaling dimension 1/16. We investigate some  properties of the interface operator including its expansion near the 
IR fixed point. 

The second quantity we consider is the effective Hamiltonian which we construct systematically both near the UV and IR fixed 
points.  While both Hamiltonians describe small energy states they have different nature. For the UV effective Hamiltonian 
we adopt the mode  truncation regularisation of \cite{Toth} and the effective interaction terms describe the low energy effects 
of the omitted high energy modes. These interaction terms are given by irrelevant operators with couplings that depend on the 
boundary magnetic field $h$ and the truncation parameter.  Such effective Hamiltonians are well understood and 
we work it out for the present model  to see how the Schrieffer-Wolff method \cite{SW} (see \cite{SWreview} for a 
review) works in this simple setting and to prepare ground 
for the infrared effective Hamiltonian which is our main interest. 

Usually in QFT when we study a flow ending up in the infrared at a non-trivial  fixed point we do not go beyond guessing the leading irrelevant 
operator  along which the theory arrives to the fixed point. But even a perturbation by a 
single irrelevant operator requires infinitely many counter terms. Furthermore irrelevant perturbations can  lead to unstable 
or non-unitary theories depending on the sign of the perturbation or when the value of the coupling is large enough. The subleading terms 
are necessary to keep the effective theory unitary and well defined at a range of scales near the fixed point. 
The leading irrelevant terms in the effective actions near infrared fixed points were studied for 
integrable RG flows in two dimensions  in \cite{AlZ1}, \cite{AlZ2}, \cite{KM},   \cite{Berkovich}, \cite{Feverati_etal}. 
For the boundary magnetic field model the leading irrelevant perturbation was identified in \cite{Toth}.  It is 
proportional to the bulk stress-energy tensor restricted to the boundary. 
In this paper we go beyond \cite{Toth} and show how one can systematically generate the subleading terms in the effective Hamiltonian 
using the Schrieffer-Wolf method. We calculate explicitly some of the subleading terms, they are presented in (\ref{heffR}). 
The two themes --  the RG interface and the effective infrared theory, get interwoven when we discuss the expansion of the 
RG interface in the far infrared. 

The main body of the paper is organised as follows. 
In section \ref{Btr} we explain how the model is solved using a Bogolyubov transformation. That includes both the spectrum and the 
vacuum. In section \ref{sec_interface} we explicitly construct the interface operator between the unperturbed and perturbed models 
and analyse its asymptotic behaviour. In section  \ref{Modtr_sec} the spectrum of the mode truncated theory is discussed along with generalisations 
of the model which include switching on perturbations by the stress-energy tensor. In section \ref{UV_sec} we discuss the effective 
Hamiltonian near the UV fixed point. We first derive it using the exact solution for stress-energy tensor perturbation described in 
section  \ref{Modtr_sec}. After that we introduce the Schrieffer-Wolf method and apply it to the same problem. In section 
\ref{IR_sec} we use the Bogolyubov transformation and a version of Schrieffer-Wolf transformations to show how one can 
systematically generate terms in the effective Hamiltonian near the infrared fixed point, explicitly computing some of the 
subleading terms. In section \ref{discussion_sec} we conclude pointing at some questions that would be 
interesting to address. Two appendices contain some calculational details.

\section{Solution via Bogolyubov transformation} \label{Btr}
\setcounter{equation}{0}
As discussed in \cite{GZ} for conformal field theory on a half plane the stress-energy tensor conservation equations 
are 
\bea
&& \partial_{\bar z} T = 0 \, , \qquad \partial_{ z} \bar T = 0 \, , \nonumber \\
&& T(x) - \bar T(x) = -2\pi i \frac{d \theta(x)}{dx}  
\eea
where as usual $T= -2\pi T_{zz}$, $\bar T = -2\pi T_{\bar z\bar z}$ are rescaled components of the stress-energy tensor 
while $\theta(x)$ is the single boundary component. The Hamiltonian for translations along the boundary can then be written as 
\be\label{Hgen}
H = \frac{1}{2\pi} \int\limits_{0}^{\infty} [T(x+iy)  + \bar T(x-iy)] dy + \theta(x)  \, .
\ee 
For a perturbation of conformal boundary condition described by a Euclidean action\footnote{We adopt a convention in which  correlation 
functions are deformed by inserting $e^{-S}$.} 
\be
S= S_{0} + \lambda \int\limits_{-\infty}^{\infty}\!\! \phi(x) 
\ee
with $\phi(x)$ being some boundary operator, one has $\theta(x)=\beta(\lambda)\phi^{\rm ren}(x) $ where $\beta(\lambda)$ is 
the beta function for the boundary coupling $\lambda$ and $\phi^{\rm ren}(x) $ stands for the renormalised boundary operator. 
In the absence of resonances the beta function is linear and one has \cite{GZ}
\be \label{theta}
\theta(x) = \lambda(\Delta - 1) \phi^{\rm ren}(x)
\ee
where $\Delta$ is the conformal dimension of  $\phi(x)$. 

To study our model using the hamiltonian formalism we put it on a strip. This gives us a discrete Hilbert space and 
regulates infrared divergences. 
Let the strip have width $L$ and coordinates $\tau \in {\mathbb R}$, $\sigma\in [0,L]$ along and across the strip respectively. 
The strip is related to the upper half plane by 
 a conformal 
transformation $w=\frac{L}{\pi}{\rm Log}( z)$ where $w=\tau + i\sigma$. The free boundary condition on both ends of the strip is 
described by 
\be
\psi(w)|_{\sigma = 0} = \bar \psi(\bar w)|_{\sigma = 0} \, , \qquad 
\psi(w)|_{\sigma = L}=-\bar \psi(\bar w)|_{\sigma = L}\, .
\ee 
This gives the following standard mode expansions for the fermion fields on the strip 
\bea
&& \psi(w) = \sqrt{\frac{\pi}{L}}\sum_{k=-\infty}^{\infty} e^{-\frac{(k+1/2)\pi}{L}(\tau + i\sigma)} a_{k+1/2} \, , \nonumber \\
&& \bar \psi(w) = \sqrt{\frac{\pi}{L}}\sum_{k=-\infty}^{\infty} e^{-\frac{(k+1/2)\pi}{L}(\tau - i\sigma)} a_{k+1/2} \, .
\eea
The corresponding creation operators are $a_{k+1/2}^{\dagger}=a_{-(k+1/2)}, k\ge 0$ that together with the annihilation operators 
$a_{k+1/2}, k\ge 0$ furnish a representation of the Neveu-Schwarz (NS) algebra. The Hamiltonian that generates translations 
along the strip is given by 
\be
H_{0} = \frac{\pi}{L}(L_{0} - \frac{c}{24}) = \frac{\pi}{L}\left( \sum_{k=0}^{\infty}(k+\frac{1}{2})a^{\dagger}_{k+1/2}a_{k+1/2} - \frac{1}{48}\right) \, .
\ee  
As we have a finite length system it no longer exhibits a degenerate vacuum. The unique vacuum $|0\rangle$ satisfies 
$a_{k+1/2}|0\rangle = 0$. The boundary fermion operator $a$ mixes the states with even and odd number of creation operators so that 
the physical state space ${\cal H}_{L}$ is spanned by the orthonormal basis 
\be \label{UV_basis}
a^{\dagger}_{k_1+1/2}\dots a^{\dagger}_{k_{N}+1/2}|0\rangle \, , \enspace \mbox{N--even} \, , \quad 
a^{\dagger}_{k_1+1/2}\dots a^{\dagger}_{k_{N}+1/2}|a\rangle \, , \enspace \mbox{N--odd} \, .
\ee
It is also convenient to consider an extended Fock space ${\cal F}_{L}$ spanned by vectors of the above type with 
parity conditions on the numbers of oscillators disregarded. We have ${\cal F}_{L}={\cal H}_{L}\oplus {\cal H}_{L}'$ where 
$ {\cal H}_{L}'$ is spanned by vectors 
\be \label{UV_basis2}
a^{\dagger}_{k_1+1/2}\dots a^{\dagger}_{k_{N}+1/2}|0\rangle \, , \enspace \mbox{N--odd} \, , \quad 
a^{\dagger}_{k_1+1/2}\dots a^{\dagger}_{k_{N}+1/2}|a\rangle \, , \enspace \mbox{N--even} \, .
\ee
The operator $a$ can  be defined on ${\cal F}_{L}$ by requiring that it 
anti-commutes  with all $a^{\dagger}_{k+1/2}$, $a_{k+1/2}$ and satisfies $a^2 = 1$, and $a|0\rangle =|a\rangle$.

We can also introduce a boundary fermion operator $a'$ (on ${\cal F}_{L}$) that leaves on the $\sigma=L$ end of the boundary. 
It satisfies $(a')^2=1$, anti-commutes with 
$a$ and all operators $a^{\dagger}_{k+1/2}$, $a_{k+1/2}$ and can be chosen to act
\be
a'|0\rangle = i|a\rangle \, , \qquad a'|a\rangle = -i|0\rangle \, .
\ee
We further form operators 
\be
A=\frac{1}{2}(a+ia') \, , \qquad A^{\dagger} = \frac{1}{2}(a-ia')
\ee
that satisfy the canonical anti-commutation relations
\be
\{A,A\}=\{A^{\dagger},A^{\dagger}\} = 0 \, , \qquad \{A, A^{\dagger}\} =1\, .
\ee
We note that 
\be
A|0\rangle = 0 \, , \qquad A^{\dagger}|0\rangle =|a\rangle \, . 
\ee
The space ${\cal F}_{L}$ is then the standard Fock space for the canonical  oscillator algebra spanned by 
$\{A,A^{\dagger},a_{k+1/2},a^{\dagger}_{k+1/2}|k=0,1,\dots\}$ with the Fock vacuum $|0\rangle$. The physical 
state space ${\cal H}_{L}$ is preserved by products of even numbers of  operators from this oscillator algebra.

Perturbing the model by the boundary magnetic field $h$ on the $\sigma=0$ end of the strip is described  on a half plane 
by a  Hamiltonian of the form (\ref{Hgen}), (\ref{theta}) with $\theta(x)$ proportional to $\sigma_{B}(x)$. 
On the strip the perturbed Hamiltonian 
 reads
\be
H_{h} = H_{0} + ih\sqrt{\frac{\pi}{L}}a\sum_{k=0}^{\infty}(a^{\dagger}_{k+1/2} + a_{k+1/2})  
\ee
that is defined on ${\cal H}_{L}$.  It is convenient to introduce a dimensionless Hamiltonian 
\be \label{halpha}
h_{\alpha} \equiv \frac{L}{\pi} H_{h} =  \sum_{k=0}^{\infty}(k+\frac{1}{2})a^{\dagger}_{k+1/2}a_{k+1/2} - \frac{1}{48} + i\alpha a \sum_{k=0}^{\infty}(a^{\dagger}_{k+1/2} + a_{k+1/2})
\ee
where 
\be
\alpha = h \sqrt{\frac{L}{\pi}}
\ee
is a dimensionless coupling. 

The RG flow sends the coupling $\alpha$ to infinity so that asymptotically the model approaches a non-trivial infrared fixed point described by the 
fixed spin boundary condition at $\sigma=0$ (the sign of the spin is fixed by the sign of the coupling $\alpha$). The infrared fixed point (keeping the spectator boundary condition at $\sigma=L$ always to be free) is 
described by Ramond oscillators $b^{\dagger}_{n}$, $b_{n}$, where $n>0$ is an integer and $b_{0}$ is the zero mode. The corresponding 
dimensionless Hamiltonian is 
\be\label{h_inf}
h_{\infty} = \sum_{n=1}^{\infty}nb^{\dagger}_{n}b_{n} + \frac{1}{24}
\ee
where the $c/24$ Casimir energy is shifted due to the $1/16$ weight of the $\sigma$-representation of the Virasoro algebra. 

We will solve the perturbed model  (\ref{halpha}) by a Bogolyubov transformation that in particular will establish a precise connection between the 
$a^{\dagger}_{k+1/2}$ and $b^{\dagger}_{n}$ oscillators. 

\subsection{The spectrum} \label{spectrum_sec}
The Hamiltonian (\ref{halpha}) is quadratic and can be diagonalised by means of a Bogolyubov transformation. To find the latter we 
write a linear ansatz
\be \label{B1}
b_{\omega} =  \sum_{k=0}^{\infty} (A_{\omega, k}a^{\dagger}_{k+1/2} + B_{\omega, k}a_{k+1/2}) + \frac{a}{f(\omega)}\, ,
\ee
subject to the equation 
\be \label{com_eq}
[h_{\alpha}, b_{\omega}] = \omega b_{\omega} \, .
\ee
The system of linear equations on the coefficients $A_{\omega, k}$, $B_{\omega, k}$, $f(\omega)$ admits a non-trivial solution 
provided $\omega$ satisfies the following transcendental equation
\be \label{spec}
 \tan(\pi \omega) = - \frac{\omega}{2\pi \alpha^2} \, ,  \quad \alpha \ne 0 \, .
\ee
It has a zero mode solution $\omega=0$ as well as an infinite number of solutions $\pm \omega_{n}$, $n=1,2,3,\dots $  with $\omega_{n+1}>\omega_{n}>0$.
The positive $\omega$ solutions $b_{\omega}$ are creation operators and the negative $\omega$ ones are the conjugated annihilation operators. 
The values $\omega_{p}$ give excitation energies above the vacuum. We will formally demonstrate this below but before that we discuss 
some general features of the solutions to (\ref{spec}) and develop further the Bogolyubov transformation.

Equation (\ref{spec}) is quite well known in the literature and in fact there are explicit integral formulas available for the  values $\omega_{p}$ derived via a suitable Riemann factorisation 
problem \cite{spec_exact}.  Most of the general features of the spectrum however can be easily obtained directly from equation (\ref{spec}) as well as  the initial pieces of power  series expansions 
near $\alpha=0$ and $\alpha=\infty$.  As we increase $\alpha$ from $\alpha=0$ towards $\alpha=\infty$ each value $\omega_{n}$  monotonically 
interpolates between $\omega_{n} = n-1/2$ at $\alpha=0$ and $\omega_{n}=n$ at $\alpha=\infty$. This corresponds to a flow between the 
Neveu-Schwarz  and Ramond algebras describing the free and fixed boundary conditions at the UV and IR ends of the RG flow.  
For  $n\gg \alpha^2$ we have 
\be \label{large_n}
 \omega_{n} = n-1/2 + \frac{2\alpha^2}{n-1/2} 
+ {\cal O}\left( \left( \frac{n}{\alpha^2}\right)^2\right) 
\ee
that means that above the energy scale $\alpha^2$ set by the coupling $h$ we recover the UV CFT spectrum of conformal 
dimensions.

Coming back to the linear transformation (\ref{B1}) we further choose a set of canonically normalised solutions  
\bea \label{Bog1}
b^{\dagger}_{\alpha, n} &=&\sum_{k=0}^{\infty}  (A_{n, k}a^{\dagger}_{k+1/2} + B_{n, k}a_{k+1/2}) + \frac{a}{f_{n}}\, , \nonumber \\
b_{\alpha, n} &=& \sum_{k=0}^{\infty}  (A^{*}_{n, k}a_{k+1/2} + B^{*}_{n, k}a^{\dagger}_{k+1/2}) + \frac{a}{f_{n}}
\eea
with 
\be
 A_{n,k}= \frac{2i\alpha}{(k+1/2-\omega_{n})f_{n}} \, , \qquad 
B_{n,k}= \frac{-2i\alpha}{(k+1/2+\omega_{n})f_{n}}\, , 
\ee
\be
 f_{n}  = \sqrt{2 + 4\alpha^2\pi^2 + \frac{\omega_{n}^2}{\alpha^2}} \, .
\ee
 The zero mode is chosen to be
\be\label{Bog2}
b_{\alpha, 0} =  \sum_{k=0}^{\infty} (A_{0, k}a^{\dagger}_{k+1/2} + B_{0, k}a_{k+1/2}) + \frac{a}{f_{0}}
\ee
with 
\be
A_{0,k}=\frac{2i\alpha}{(k+1/2)f_{0}}\, , \quad B_{0,k}= \frac{-2i\alpha}{(k+1/2)f_{0}} \, , \quad f_{0}= \sqrt{2 + 4\alpha^2\pi^2 } \, . 
\ee
These solutions satisfy the canonical anti-commutation relations 
\be\label{CAR}
\{b^{\dagger}_{\alpha,n}, b_{\alpha, m}\} = \delta_{n,m}\, , \enspace \{b^{\dagger}_{\alpha,n}, b^{\dagger}_{\alpha,m}\} =  
\{b_{\alpha,n}, b_{\alpha, m}\} = 0 \, , n,m>0 \, , 
\ee
\be
\{ b_{\alpha, 0}, b_{\alpha, 0}\} = 2b_{\alpha,0}^2=1\, , \quad \{b_{\alpha,0},b_{\alpha, n} \}=\{b_{\alpha,0},b^{\dagger}_{\alpha, n} \}=0\, .
\ee
The phases of these solutions are chosen based on computational convenience. With this choice we have 
\be
\lim_{\alpha \to 0} b_{\alpha, 0} = \frac{a}{\sqrt{2}} \, , \quad \lim_{\alpha \to 0} b^{\dagger}_{\alpha, n} = i{\rm sign}(\alpha) a^{\dagger}_{n-1/2} 
\quad \lim_{\alpha \to 0} b_{\alpha, n} =-i{\rm sign}(\alpha) a_{n-1/2} \, .
\ee
We further define operators 
\be
B_{\alpha}=\frac{1}{2}(\sqrt{2}b_{\alpha, 0} + ia') \, , \qquad B_{\alpha}^{\dagger}=\frac{1}{2}(\sqrt{2}b_{\alpha, 0} - ia')
\ee
that anti-commute with all operators $b^{\dagger}_{n}, b_{n}$, $n\in {\mathbb N}$ and satisfy 
\be
\{B_{\alpha},B_{\alpha}\}=\{B^{\dagger}_{\alpha}, B^{\dagger}_{\alpha}\}=0\, , \qquad \{B_{\alpha}, B^{\dagger}_{\alpha}\}=1 \, .
\ee
The equations (\ref{Bog1}), (\ref{Bog2})  then define a linear canonical transformation from the fermionic oscillator algebra ${\cal A}$ spanned by 
$\{A,A^{\dagger},a_{k+1/2},a^{\dagger}_{k+1/2}|k=0,1,\dots\}$ and the fermionic algebra ${\cal B}$ spanned by 
$\{B_{\alpha},B^{\dagger}_{\alpha},b_{\alpha, n},b^{\dagger}_{\alpha, n}|n=1,2,\dots \}$. 

Next we show that the canonical transformation at hand 
is realised by a unitary operator that acts in ${\cal F}_{L}$ and preserves the subspaces ${\cal H}_{L}$, ${\cal H}_{L}'$. 
(It thus deserves to be called a Bogolyubov transformation.)

First we remind the reader some general facts (see e.g. \cite{Berezin}). 
A general canonical transformation acting on oscillator algebra spanned by  $a^{\dagger}_{i}, 
a_{i}$ with $i$ from some discrete index set, can be written as 
\be \label{Bog_gen}
a_{i}'=\sum_{j}(\Phi_{ij}a_{j} + \Psi_{ij}a^{\dagger}_{j})\, , \qquad (a_{i}')^{\dagger}=\sum_{j}(\Phi_{ij}^{*}a^{\dagger}_{j} + \Psi_{ij}^{*}a_{j}) \, .
\ee
A unitary operator $U$ such that 
\be
a'_{i}=Ua_{i}U^{-1} \, , \qquad (a'_{i})^{\dagger}=Ua_{i}^{\dagger}U^{-1} 
\ee
exists if and only if the operator $\Psi_{ij}$ is Hilbert Schmidt that is if 
\be 
{\rm Tr}\Psi^{\dagger}\Psi = \sum_{i,j}|\Psi_{ij}|^2 < \infty \, .
\ee
In the case at hand we calculate from (\ref{Bog1}), (\ref{Bog2}) and the definitions of $A^{\dagger}, A, B_{\alpha}, B^{\dagger}_{\alpha}$
\bea
&& {\rm Tr} {\Psi}^{\dagger}{\Psi} =\sum_{n=0}^{\infty}\sum_{k=0}^{\infty} |B_{n,k}|^2 +  \frac{1}{2}\sum_{k=0}^{\infty}|B_{0,k}|^2+ \sum_{n}^{\infty} \frac{1}{|f_{n}|^2} 
\nonumber \\
&& = 
\frac{\pi^2\alpha^2}{f_{0}^2}+ \sum_{n=1}^{\infty}\frac{\alpha^2\psi'(1/2+\omega_{n})+1}{f_{n}^2} < \infty 
\eea
because of (\ref{large_n}). 
The above inequality is true for every finite value of $\alpha$. 
In section \ref{sec_interface} we will explicitly construct the corresponding unitary operator. Since this operator preserves the numbers of 
oscillators it preserves the physical space  ${\cal H}_{L}$ and the complimentary space ${\cal H}_{L}'$.
The Fock vacuum state $|0\rangle_{\alpha}$ of the algebra generated by 
$\{B_{\alpha},B^{\dagger}_{\alpha},b_{\alpha, n},b^{\dagger}_{\alpha, n}|n=1,2,\dots \}$ 
satisfies $B_{\alpha}|0\rangle=b_{\alpha, n}|0\rangle=0$, $n=1,2, \dots$. As is well known in the theory of Bogolyubov transformations the 
vacuum $|0\rangle$ can be written as an exponential of an operator quadratic in the creation operators $a^{\dagger}_{k+1/2}$, $A^{\dagger}$ 
and thus belongs to ${\cal H}_{L}$.

We now turn to the Hamiltonian which being rewritten in terms of the new set of oscillators is 
\be \label{ham2}
h_{\alpha} = \sum_{n=1}^{\infty} \omega_{n} b^{\dagger}_{\alpha, n}b_{\alpha, n} + {\cal E}_{\alpha}
\ee
where\footnote{This expression can be formally derived by substituting the linear transformations (\ref{Bog1})  into 
$\sum  \omega_{n} b^{\dagger}_{\alpha, n}b_{\alpha, n}$, normal ordering the expression with respect to the oscillators $a^{\dagger}_{k+1/2}$, 
$a_{k+1/2}$ and then using the standard bilinear relations 
for the coefficients of the Bogolyubov transformation stemming from the canonical anti-commutation relations (\ref{CAR}). Also some regularisation of the sums should be assumed in these manipulations.} 
\be
{\cal E}_{\alpha} = -\frac{1}{48} + \frac{1}{2}\sum_{n=1}^{\infty} ( n-1/2 - \omega_{n}) 
\ee
is the new vacuum energy. It follows from (\ref{large_n}) that this sum  diverges as a harmonic series. This is the only divergence in the theory 
that is removed by a constant counterterm. In the Lagrangian description this divergence shows up as a one loop divergence in 
 the integral of  the two point function $\langle a(\tau)\psi(\tau) a(\tau')\psi(\tau') \rangle$ that has a $1/|\tau-\tau'|$ singularity. 
Using  contour  integration  the subtracted energy can be written as 
\be
{\cal E}_{\alpha}^{\rm ren} = -\frac{1}{48} - \frac{1}{2\pi}\int\limits_{0}^{\infty} \!\! dx \, \ln\left( \frac{x + 2\pi \alpha^2 \tanh(\pi x)}{x+2\pi\alpha^2} \right) 
+  \alpha^2 \ln(\alpha^2 \mu) 
\ee
where $\mu$ is a subtraction point. In the limit $\alpha \to \infty$ the value of the integral term becomes $1/16$ that is the lowest conformal 
weight for fixed-free boundary conditions on a strip.

Evidently the states minimising the energy of (\ref{ham2}) considered as an operator on ${\cal F}_{L}$ are linear combinations of the Fock vacuum 
$|0\rangle_{\alpha}$ and the state $B^{\dagger}_{\alpha}|0\rangle_{\alpha}$. Since as explained before $|0\rangle_{\alpha}\in {\cal H}_{L}$ 
it follows that  $B^{\dagger}_{\alpha}|0\rangle_{\alpha}\in {\cal H}_{L}'$ and therefore $|0\rangle_{\alpha}$ is the unique vacuum  
of the Hamiltonian 
$h_{\alpha}$ that lies in the physical state space ${\cal H}_{L}$. All excited energy eigenstates are obtained by acting on 
$|0\rangle_{\alpha}$ by an even number of creation oscillators $B^{\dagger}_{\alpha}, b^{\dagger}_{\alpha, n}$. Since $B^{\dagger}_{\alpha}$ 
commutes with $h_{\alpha}$ the energy spectrum is given by a sum of ${\cal E}_{\alpha}$ and a number of values $\omega_{p}$
without repetition. This spectrum can be easily extrapolated to $\alpha=\infty$ that clearly coincides with the free fermion one that 
describes (\ref{h_inf}).

\subsection{Some further formal aspects}

It is straightforward to find that the inverse transformation to (\ref{Bog1}), (\ref{Bog2})  is 
\bea\label{inverse1}
a_{k+1/2}^{\dagger} &=& \sum_{n=1}^{\infty} (A^{*}_{n, k}b^{\dagger}_{\alpha,n} +B_{n, k}b_{\alpha,n}) + A^{*}_{0,k}b_{0} \, , \nonumber \\
a_{k+1/2} &=& \sum_{n=1}^{\infty} (A_{n, k}b_{\alpha,n} +B_{n, k}^{*}b_{\alpha,n}^{\dagger}) + A_{0,k}b_{0} \, ,
\eea
\be \label{inverse2}
a = 2\Bigl[ \sum_{n=1}^{\infty} \frac{b^{\dagger}_{\alpha,n} + b_{\alpha,n}}{f_{n}} + \frac{b_{0}}{f_{0}}\Bigr] \, . 
\ee

The linear transformation (\ref{Bog1}), (\ref{Bog2}) can be continued to the limit $\alpha=\infty$ where all coefficients tend to finite limits. 
Explicitly we get a linear transformation 
\bea \label{Asympt_transf}
b^{\dagger}_{ n} &=&\sum_{k=0}^{\infty}  (A^{\infty}_{n, k}a^{\dagger}_{k+1/2} + B_{n, k}^{\infty}a_{k+1/2}) \, , \nonumber \\
b_{ n} &=& \sum_{k=0}^{\infty}  (( A^{\infty}_{n, k})^{*}a_{k+1/2} + (B^{\infty}_{n, k})^{*}a^{\dagger}_{k+1/2}) \, , \nonumber \\
b_{0}&=& \sum_{k=0}^{\infty}  (A^{\infty}_{0, k}a^{\dagger}_{k+1/2} + B_{0, k}^{\infty}a_{k+1/2})
\eea
with 
\be \label{limitAB1}
 A_{n,k}^{\infty}=  \frac{i}{\pi(k+1/2- n)} \, , \qquad 
B_{n,k}^{\infty}=  \frac{-i}{\pi(k+1/2+n)} \, ,
\ee
\be\label{limitAB2}
 A_{0,k}^{\infty}=  \frac{i}{\pi(k+1/2)} \, , \qquad 
B_{0,k}^{\infty}=  \frac{-i}{\pi(k+1/2)} \,  .
\ee
We see that the boundary fermion $a$ drops out of the linear transformation and 
thus can no longer  be obtained as a linear combination of $b^{\dagger}_{n}$, $b_{ n}$, $b_{0}$. (If we take the limit $\alpha\to \infty$ 
mode-wise in (\ref{inverse2}) we obtain zero on the right hand side.)
It is straightforward to check though that (\ref{Asympt_transf}) still defines a canonical transformation between the set of modes 
$\{ a^{\dagger}_{k+1/2}, a_{k+1/2}| 
k\in {\mathbb N}\}$ and the set $\{b_{0},b^{\dagger}_{n}, b_{n}|n\in {\mathbb N}\}$.  The inverse transformation is obtained by substituting 
(\ref{limitAB1}), (\ref{limitAB2}) into (\ref{inverse1}). This transformation however is no longer a Bogolyubov transformation\footnote{In terminology of \cite{Berezin} 
we have here an improper canonical transformation. } 
as 
\be
\sum_{n=1}^{\infty}\sum_{k=0}^{\infty} |B_{n,k}^{\infty}|^2 
\ee
diverges logarithmically  (as a harmonic series $\sum_{n=1}^{\infty} \frac{1}{n}$ ).  This fact has an interesting interpretation in terms of 
the RG interface operator that we discuss in section \ref{sec_interface}. The modes $\{b_{0},b^{\dagger}_{n}, b_{n}|n\in {\mathbb N}\}$ 
are associated with the Ramond free fermion description of the IR fixed point with Hamiltonian (\ref{h_inf}).

We note in passing that
we  also derived a Bogolyubov transformation between the oscillators of two deformed theories with couplings $\alpha_1$ and $\alpha_2$. 
Taking $\alpha_2$ to infinity we find a transformation 
that is no longer a canonical linear transformation.

\subsection{The vacuum}
In this section we will construct explicitly the vacuum state $|0\rangle_{\alpha}$ in terms of the UV CFT oscillators 
$a^{\dagger}_{k+1/2}$ and the boundary fermion $a$. For a general Bogolyubov transformation (\ref{Bog_gen}) 
the Fock space vacuum annihilated by all of the annihilation oscillators $a'_{i}$ is given by 
\be
|0\rangle'=\uptheta\, {\rm det}\! \left( \Phi \Phi^{\dagger}\right)^{1/4} \exp\left[  -\frac{1}{2}a^{\dagger}_{i} (\Phi^{-1}\Psi)_{ij}a^{\dagger}_{j} 
\right] |0\rangle \, 
\ee
where $\uptheta$ is a phase that can be chosen arbitrarily. For the Bogolyubov transformation at hand the matrix 
$\Phi_{ij}$ is an infinite-dimensional matrix that does not have any simple structure that would allow us to invert 
it explicitly. To obtain an explicit form of the vacuum we proceed in a roundabout way. 

Writing an ansatz 
\be\label{vac_ans}
|0\rangle_{\alpha} = {\cal N}_{\alpha} \exp\left[ \frac{1}{2}\sum_{p,r=0}^{n_c}a^{\dagger}_{p+1/2}M_{pr} a_{p+1/2}^{\dagger} \right] 
\left( |0\rangle + i\sum_{l=0}^{n_c}d_{l}a^{\dagger}_{l+1/2}|a\rangle \right) 
\ee
we require that it is an eigenvector  of the perturbed Hamiltonian (\ref{halpha}) with an eigenvalue ${\cal E}_{0}$. As ${\cal E}_{0}$ is expected to be 
divergent we 
regularise all sums over mode numbers by truncating them at some value $n_{c}\gg 1$ to be later taken to infinity. 
Direct calculation yields the following set of equations 
\bea\label{Mdd}
&& (p+r+1)M_{pr} -\alpha( d_{r}-d_{p} + \sum_{k=0}^{n_c} [M_{kp}d_{r} -M_{kr}d_{p}])=0 \, , \nonumber \\
&& d_{p}(p+1/2) - \alpha \left(1 + \sum_{k=0}^{n_c}M_{kp}\right) = {\cal E}_{0}d_{p} \, , \nonumber \\
&& -\alpha \sum_{k=0}^{n_c} d_{k} = {\cal E}_{0}\, .
\eea
Combining the first two equations we obtain 
\be \label{M}
M_{pr}=d_{p}d_{r}\left(\frac{p-r}{r+p+1}\right)  \qquad r,p=0,1,2,\dots 
\ee
Substituting (\ref{M}) and the third equation in (\ref{Mdd}) into the second equation in (\ref{Mdd}) we get the following 
system of quadratic equations on the vector $d_{p}$
\be \label{d_recurrence}
(p+1/2)d_{p}\left( 1 + 2\alpha \sum_{k=0}^{\infty}\frac{d_{k}}{p+k+1} \right) = \alpha \qquad p=0,1,2,\dots 
\ee
where we already took the limit $n_{c}\to \infty$. If we solve (\ref{d_recurrence}) then using (\ref{M}), (\ref{vac_ans}) we obtain 
the vacuum vector. 

Next we define a generating function 
\be \label{gen_f}
f(t) = \sum_{k=0}^{\infty} d_{k}e^{-(k+1/2)t} \, , \quad t> 0\, .
\ee
We see from (\ref{d_recurrence}) that 
for $p\to \infty$, neglecting the sum, we have $d_{p} \to \frac{\alpha}{p+1/2}$ so that the series in (\ref{gen_f}) 
converges uniformly for all $t>0$.

In terms of the generating function (\ref{gen_f}) the system of equations (\ref{d_recurrence}) takes the form of an integral equation
\be \label{int_eq}
f(t) + 2\alpha \int\limits_{0}^{\infty}dt'\, f(t+t')f(t') = \alpha G(t) \qquad t\ge 0
\ee
where 
\be
G(t) = \sum_{l=0}^{\infty} \frac{e^{-(l+1/2)t}}{l+1/2} = \ln\left( \frac{1+ e^{-t/2}}{1-e^{-t/2}} \right) \, .
\ee
To extend the above integral equation to the entire real line we define 
\be
f_{+}(t) =  \left \{
\begin{array}{l@{\enspace}l}
f(t) \, ,  &t> 0
\\[1ex]
0\, , & t\le 0
\end{array}
\right .  \, , 
\qquad f_{-}(t) = \left \{
\begin{array}{l@{\enspace}l}
0\, ,  &t\ge  0
\\[1ex]
f(-t) \, , &t< 0
\end{array}
\right .
\ee
and 
\be
G_{+}(t) =  \left \{
\begin{array}{l@{\enspace}l}
G(t) \, ,  &t> 0
\\[1ex]
0\, , & t\le 0
\end{array}
\right .  \, , 
\qquad G_{-}(t) = \left \{
\begin{array}{l@{\enspace}l}
0\, ,  &t\ge  0
\\[1ex]
G(-t) \, , &t< 0
\end{array}
\right . \, .
\ee
 In terms of these functions we recast equation (\ref{int_eq}) as 
 \be \label{int_eq2}
 f_{+}(t) + f_{-}(t) + 2\alpha \int\limits_{0}^{\infty}dt'\, f_{+}(t+t')f_{+}(t') = \alpha(G_{+}(t) + G_{-}(t)) \quad t\in {\mathbb R}
 \ee
 
We next define the Fourier transforms 
\be
\tilde f_{\pm}(\omega) = \frac{1}{2\pi}\int\limits_{-\infty}^{\infty}dt\, f_{\pm}(t) e^{-i\omega t} \, , \quad 
\tilde G_{\pm}(\omega) = \frac{1}{2\pi}\int\limits_{-\infty}^{\infty}dt\, G_{\pm}(t) e^{-i\omega t}\, .
\ee
The function $\tilde f_{+}(\omega)$ analytically extended to the complex plane is holomorphic in the lower half plane ${\rm Im}(\omega)<0$.
Also, since  $\tilde f_{-}(\omega) = \tilde f_{+}(-\omega)$  the analytically extended $\tilde f_{-}(\omega)$ is holomorphic in the upper half plane. Fourier transforming (\ref{int_eq2}) we obtain 
\be
\tilde f_{+}(\omega) + \tilde f_{+}(-\omega) + 4\pi\alpha\tilde f_{+}(\omega)\tilde f_{+}(-\omega) = \alpha(\tilde G_{+}(\omega) + \tilde G_{+}(-\omega)) 
= \frac{\alpha\tanh(\pi \omega)}{2\omega}\, .
\ee
Finally, defining 
\be 
\Phi(\omega) = 1 + 4\pi\alpha \tilde f_{+}(\omega) 
\ee
we rewrite the last equation as  a holomorphic factorisation equation 
\be\label{factorisation}
\Phi(\omega) \Phi(-\omega) = 1 + \frac{2\pi\alpha^2\tanh(\pi\omega)}{\omega} \, .
\ee
This can be solved by standard techniques. Relegating the details to appendix A we present here an integral formula 
for $d_{k}$ that follows from the solution to (\ref{factorisation}) 
\be \label{dk}
d_{k} = \frac{\Gamma(k+1/2)}{\sqrt{2\pi}\Gamma(k+1)}\exp\left[    -\frac{(k+1/2)}{2\pi}\int\limits_{-\infty}^{\infty}\frac{dx}{x^2+(k+1/2)^2} 
\ln\left( 1 + \frac{x}{2\pi\alpha^2\tanh(\pi x)}\right)
\right]
\ee

\section{The RG interface operator}\label{sec_interface}
\setcounter{equation}{0}
\subsection{Construction of the operator}

In the operator quantisation we can associate with every interface between two QFTs a pairing of states in the two theories (see 
\cite{TCSA_Ising} for a general discussion). For the boundary RG flows on a strip such a pairing  can be defined by means of an RG  interface operator 
\be
\hat \psi_{0,\lambda}: {\cal H}^{\lambda} \to {\cal H}^{0}
\ee
that acts from the state space of the UV BCFT -- ${\cal H}^{0}$ to the perturbed theory state space --
 ${\cal H}_{\lambda}$. We can choose  bases of energy eigenstates: $\{ |E_{i}\rangle_{0} \} $ in ${\cal H}^{0}$ and 
 $\{|E_{j}\rangle_{\lambda}\}$ in  ${\cal H}^{\lambda}$. Perturbation theory allows us to express 
 the eigenvector of the perturbed theory as vectors in  ${\cal H}^{0}$ (or rather in the dual space due to Haag's theorem). 
In superrenormalisable theories while the norms of the perturbed 
eigenvectors defined in the unperturbed state space are divergent, the ratio's of components are finite. Thus up to normalisation we can define an embedding 
of the eigenvectors that specifies  the action of the interface operator:
\be \label{bases}
\hat \psi_{0,\lambda} |E_{j}\rangle_{\lambda}=\sum_{ji}C_{ji}|E_{i}\rangle_{0}
\ee
where $C_{ji}$ are  numbers.
The coefficients $C_{ji}$, defined up to an overall normalisation of $|E_{j}\rangle_{\lambda}$, are the pairings defined by RG interface. 
They are  matrix elements of the RG operator $\hat \psi_{ 0,\lambda} $:
\be
C_{ij} = {}_{0}\langle E_{i}|\hat \psi_{0, \lambda} |E_{j}\rangle_{\lambda} \, .
\ee
For the boundary magnetic field model considered in this paper the expansion (\ref{bases}) is obtained via the Bogolyubov transformation 
solving the perturbed QFT. The corresponding RG operator that we denote as $\hat \psi_{0,\alpha}$ intertwines the Heisenberg algebras 
${\cal A}$ and ${\cal B}$ defined in section \ref{spectrum_sec}. 

For a general Bogolyubov transformation (\ref{Bog_gen}) such an intertwiner ${\cal I}_{0,0'}$  is defined as follows. Let ${\cal F}$ be a Fock 
space representation corresponding to the canonical algebra with generators $a_{i}, a^{\dagger}_{i}$ and a vacuum $|0\rangle$ and let 
$U:  {\cal F} \to {\cal F}$ be the unitary operator effecting the Bogolyubov transformation (\ref{Bog_gen}).
Define a separate Fock space ${\cal F}'$  with vacuum $|0'\rangle$ realising a canonical algebra with generators $b_{i}, b^{\dagger}_{i}$. 
The intertwiner ${\cal I}_{0,0'}:{\cal F}'\to {\cal F}$ is then defined so that 
\be
{\cal I}_{0,0'} b_{i} = 
\sum_{j}(\Phi_{ij}a_{j} + \Psi_{ij}a^{\dagger}_{j}) {\cal I}_{0,0'}  \qquad 
{\cal I}_{0,0'} b^{\dagger}_{i} = \sum_{j}(\Phi_{ij}^{*}a^{\dagger}_{j} + \Psi_{ij}^{*}a_{j}) {\cal I}_{0,0'} \, .
\ee
Explicitly it can be written as 
\be \label{gen_int}
{\cal I}_{0,0'}=\uptheta\, {\rm det}\! \left( \Phi \Phi^{\dagger}\right)^{1/4} |0\rangle{}\langle 0'| \exp\left[  -\frac{1}{2}\left(  
b_{i} (\Psi^{*}\Phi^{-1})_{ij}b_{j} - 2a^{\dagger}_{i}\Phi^{-1}_{ij}b_{j} + 
a^{\dagger}_{i} (\Phi^{-1}\Psi)_{ij}a^{\dagger}_{j} \right) 
\right]
\ee
where all operators $a^{\dagger}_{i}$ are understood to be standing to the left of $|0\rangle$. 
The coefficients standing at monomials in the creation and annihilation operators are the same as the coefficients 
defining the symbol of the unitary operator $U$ (see \cite{Berezin} formula (5.15)). 
Given the matrix 
${\cal M}_{ij}=(\Phi^{-1}\Psi)_{ij}$ one can express the other two matrices entering (\ref{gen_int}) as 
\be \label{other_matr}
\Phi^{-1}_{ij} = {\Phi}^{\dagger}_{ij}+ \sum_{k}{\cal M}_{ik} \Psi^{\dagger}_{kj} \, , \quad 
(\Psi^{*}\Phi^{-1})_{ij} = \sum_{k} \Psi^{*}_{ik}\Phi^{\dagger}_{kj} + \sum_{kl} \Psi^{*}_{ik}{\cal M}_{kl} \Psi^{\dagger}_{lj} \, .
\ee

 We next specialise the general expression (\ref{gen_int}) to the boundary magnetic field model. As before let ${\cal F}_{L}$ be the 
Fock space corresponding to the algebra ${\cal A}$. 
Let us  introduce an independent Fock space ${\cal F}_{L}^{\alpha}$  built from the vacuum $|0\rangle_{\alpha}$ annihilated by 
$\{B_{\alpha},b_{\alpha,n}|n=1,2,\dots \}$ with the other states obtained by acting on the vacuum by the raising operators 
$\{ B^{\dagger}_{\alpha}, b^{\dagger}_{\alpha,n}|n=1,2,\dots\}$. The RG operator $\hat \psi_{0,\alpha}:{\cal F}_{L}^{\alpha} \to {\cal F}_{L}$ 
can then be worked out using (\ref{gen_int}), (\ref{other_matr}) and (\ref{M}), (\ref{dk}). Relegating the details to  appendix A
we present here the final answer
\bea \label{psi_hat1}
&& \hat \psi_{0,\alpha} = {\cal N}_{\alpha} |0\rangle{}\langle 0|_{\alpha} 
 \exp\Bigl[ \frac{1}{2}\sum_{p,r=0}^{\infty}a^{\dagger}_{p+1/2}M_{pr} a_{r+1/2}^{\dagger}  + \frac{1}{2} 
\sum_{n,m=0}^{\infty} b_{\alpha,n}  N_{nm} b_{\alpha,m} 
\nonumber \\
&& + i\sum_{n,p=0}^{\infty} a^{\dagger}_{p+1/2} O_{pn}b_{\alpha,n} \Bigr] 
\left( 1  + \sum_{n=0}^{\infty} \tilde d_{n} a b_{\alpha,n}    \right)  \left( 1 + i\sum_{l=0}^{\infty}d_{l}a^{\dagger}_{l+1/2}a  \right) 
\eea
where all operators $a^{\dagger}_{p+1/2}$ and $a$ act on $|0\rangle $ from the left e.g. 
\be
|0\rangle{}\langle 0|_{\alpha} a^{\dagger}_{l+1/2} b_{\alpha,m} a^{\dagger}_{r+1/2} b_{\alpha,n} \equiv   
 -a^{\dagger}_{l+1/2}a^{\dagger}_{r+1/2} |0\rangle{}\langle 0|_{\alpha}b_{\alpha,m}b_{\alpha,n} \, , 
\ee
the matrix $M_{pr}$ and the vector $d_{l}$ are given in (\ref{M}), (\ref{dk}), 
while 
\be
 N_{nm} = \tilde d_{n} \tilde d_{m} \left(\frac{\omega_{m}-\omega_{n}}{\omega_{m}+\omega_{n}}  \right) \, ,  \enspace m+n\ne 0\, ,  \qquad 
 N_{0,0}=0 \, , 
 \ee
 \be
 O_{pn} = d_{p}\tilde d_{n} \left(\frac{p+1/2 +\omega_{n}}{p+1/2 -\omega_{n}}  \right)\, , 
\ee
\be \label{dtwiddle}
\tilde d_{p} = \frac{\sqrt{2\pi} |\alpha|\Gamma(\omega_{p}+1/2)}{f_{p}\Gamma(\omega_{p}+1)}
\exp\left[    \frac{\omega_{p}}{2\pi}\int\limits_{-\infty}^{\infty}\frac{dx}{x^2+\omega_{p}^2} 
\ln\left( 1 + \frac{x}{2\pi\alpha^2\tanh(\pi x)}\right)\right]  
\ee
where $p=1,2,\dots $ and 
\be
\tilde d_{0} = \frac{1}{\sqrt{2}} \, .
\ee
The normalisation factor ${\cal N}_{\alpha}$ can be expressed as a determinant of a certain infinite-dimensional matrix, we will not pursue  
an explicit expression for it. It will suffice to know that it is non-vanishing for all finite values of $\alpha$ and goes to zero
 for $\alpha\to \infty$. 
Note that $\hat \psi_{0,\alpha}$ preserves the physical subspace ${\cal H}^{\alpha}_{L}\subset {\cal F}^{\alpha}_{L}$ that is built by acting 
on the Fock vacuum $|0\rangle_{\alpha}$ by even numbers of creation operators. The physical subspace is mapped onto the physical 
subspace ${\cal H}_{L}$. Since the Bogolyubov transformation is invertible there is also the inverse operator $\hat \psi_{\alpha, 0}$. 

\subsection{Further analysis} 
We next consider the $\alpha \to \infty$ limit of the RG operator $\hat \psi_{0,\alpha}$. We have $\lim_{\alpha \to \infty} \omega_{p} = p$, 
\be
d_{k} \to \frac{\Gamma(k+1/2)}{\sqrt{2\pi}\Gamma(k+1)}\, ,  \qquad \tilde d_{p} \to \frac{\Gamma(p+1/2)}{\sqrt{2\pi}\Gamma(p+1)}\equiv g(p)
\ee
where $k,p=0,1,\dots$. The normalisation factor ${\cal N}_{\alpha}$ vanishes in the $\alpha \to \infty$ limit. This
reflects the fact that the canonical  transformation  at hand stops being proper. Multiplying by a suitable function of $\alpha$  that 
diverges for $\alpha \to \infty$ we obtain in the limit a finite 
non-vanishing 
 quantity we denote ${\cal N}_{\infty}^{\rm ren}$. It is convenient to introduce 
 \be
 a^{\dagger}_{0}\equiv  i a \, 
 \ee
 and to set $g(-1/2) = 1$ 
 so that we can compactly write 
\bea
&& \hat \psi_{0,\infty} = {\cal N}_{\infty}^{\rm ren} |0\rangle{}\langle 0|_{\infty} 
 \exp\Bigl[ \frac{1}{2}\sum_{p,r=-1/2}^{\infty}a^{\dagger}_{p+1/2}M_{pr} ^{\infty} a_{r+1/2}^{\dagger}  + \frac{1}{2} 
\sum_{n,m=0}^{\infty} b_{n}  N_{nm}^{\infty} b_{m} 
\nonumber \\
&& + i\sum_{n=0}^{\infty}\sum_{p=-1/2}^{\infty} a^{\dagger}_{p+1/2} O_{pn}^{\infty}b_{n} \Bigr] 
\eea
where 
\be
M_{pr}^{\infty}=g(p)g(r)\left(\frac{p-r}{r+p+1}\right)  \, ,   \qquad 
O_{pn}^{\infty} = g(p)g(n) \left(\frac{p+1/2 +n}{p+1/2 - n}  \right) \, , 
\ee
\be
N_{n,m}^{\infty} = g(n) g(m) \left(\frac{m-n}{m+n}  \right) \, , \enspace n+m\ne 0 \, ,  \qquad N_{0,0}^{\infty}=0 \, .
\ee
and the indices $p,r$ in the sums are understood to run over the set $\{ -1/2,0,1,2,\dots \}$.

To define matrix elements of $\hat \psi_{0,\infty}$ between physical states it is convenient to introduce a basis in each physical 
space. For an orthonormal basis in ${\cal H}_{L}$ we take 
\be
 |k_{1}\dots k_{2n}\rangle_{\rm NS} = a^{\dagger}_{k_1+1/2}\dots a^{\dagger}_{k_{2n} +1/2}|0\rangle  
\ee
where  $k_{1}>k_{2}>\dots > k_{2n}$ and $k_{2n}$ is allowed to be $-1/2$. For an orthonormal basis in ${\cal H}_{L}^{\infty}$ we take 
\be
|n_{1}\dots n_{2l}\rangle_{\rm R} = b_{n_1}^{\dagger} \dots b_{n_{2l}}^{\dagger}|0\rangle_{\infty} \, , \quad n_{1}>n_{2}>\dots >n_{2l}\ge 0 \, .
\ee
We note that in standard notation for Ramond fermions the Fock vacuum $|0\rangle_{\infty}$ stands for $|\sigma\rangle$ - 
the state of lowest conformal weight $1/16$ and $b_{0}|\sigma\rangle$ stands for the disorder operator state $|\mu\rangle$. 
 
A matrix element 
\bea \label{mel}
 &&_{\rm NS}\langle k_{1}\dots k_{2s}|\hat \psi_{0,\infty} | n_1 \dots n_{2t} \rangle_{\rm R} 
 \nonumber \\
&& = {\cal N}_{\infty}^{\rm ren}\prod_{i=1}^{s}g(k_i)\prod_{j=1}^{t}g(n_j) G(k_1, \dots k_{2s}|n_1\dots n_{2t})
\eea
where $G(k_1, \dots k_{2s} |n_1\dots n_{2t})$ can be expressed as a correlator of fictitious free fermion fields $\chi(\alpha_{i}) $ 
and $\tilde \chi(\beta_{j})$ with 
\be
n_i=e^{\beta_{i}}\, , \qquad k_{j}+\frac{1}{2}=e^{\alpha_{j}}
\ee
and  two-point functions
 \be \label{prop1}
\langle \chi(\alpha_i) \chi(\alpha_{j})\rangle \equiv  -\frac{k_i-k_j}{k_i+k_j+1} =  -\tanh\left( \frac{\alpha_i - \alpha_j}{2}\right) \, , 
\ee
\be\label{prop2}
\langle \tilde \chi(\beta_{i}) \tilde \chi(\beta_{j})\rangle \equiv  \frac{n_i-n_j}{n_i+n_j} =  \tanh\left( \frac{\beta_i - \beta_j}{2}\right) \, , 
\ee
\be\label{prop3}
\langle \chi(\alpha_{l}) \tilde \chi(\beta_{j}) \rangle \equiv  i\frac{k_l+1/2 +n_j}{k_l+1/2 - n_j} =   i\coth\left( \frac{\alpha_l - \beta_j}{2}\right) \, , 
\ee
In the above  $k_i=-1/2$  corresponds to  $\alpha_{i}=-\infty$, $n_{j}=0$ corresponds to $\beta_{j}=-\infty$ and the propagator formulas are 
 obtained as limits of  (\ref{prop1}), (\ref{prop2}), (\ref{prop3}). 
The quantity  
\be
G(k_1, \dots k_{2s}|n_1\dots n_{2t}) = \langle \chi(\alpha_{2s}) \dots \chi(\alpha_{1}) \tilde \chi(\beta_{1}) \dots \tilde \chi(\beta_{2t})\rangle 
\ee
is then given as a sum over all possible contractions with the above two-point functions. 
Using a well-known identity (see e.g. \cite{YZ} formula (3.14))
\be
\langle \chi(\alpha_{2s}) \dots \chi(\alpha_{1})\rangle = \prod_{i<j}^{2s}\tanh\left(\frac{\alpha_i - \alpha_{j}}{2} \right) \, , 
\ee
which can be obtained via bosonisation, 
we obtain the following closed expression for $G$
\bea
&& G(k_1, \dots k_{2s}|n_1\dots n_{2t}) = \left(\prod_{i<j}^{2s}\frac{k_i-k_j}{k_i+k_j+1}\right) \left(\prod_{i<j}^{2t} \frac{n_i-n_j}{n_i+n_j}\right)
\nonumber \\
&& \times \left(\prod_{1\le i\le 2s; 1\le j\le 2t}  \frac{k_i+1/2 +n_j}{k_i+1/2 - n_j}\right) \, .
\eea

Formula (\ref{mel}) is to be compared with a formula for matrix elements of the primary field $\sigma$ of dimension $1/16$ 
that is a boundary condition changing field between the free and fixed boundary conditions. An explicit formula for its 
matrix elements can be obtained by taking a massless limit of formula (2.12)  in \cite{FZ} for the matrix elements of 
the bulk field $\sigma(z,\bar z)$ and separating the holomorphic factor. Adjusting for the change of OPE coefficients and 
choice of phases we find a precise match. Namely identifying the state 
$-ia^{\dagger}_{1/2}|a\rangle$ with $|\epsilon\rangle$ -- the state associated with the energy density primary, we obtain 
\be
\hat \psi_{0,\infty} = -{\cal N}_{\infty}^{\rm ren} \sigma 
\ee
where $\sigma$ is normalised so that the state it creates at infinity has norm 1 (or, equivalently, its two-point function on a half-plane 
at unit separation equals 1). While for finite $\alpha$ the interface operator $\hat \psi_{0,\alpha}$ is canonically normalised it is not clear whether there is 
such a normalisation for $\hat \psi_{0,\infty}$. 
It would be interesting to know how exactly ${\cal N}_{\alpha}$ diverges when $\alpha\to \infty$ to see if there is a canonical 
choice of ${\cal N}^{\, \rm ren}_{\infty}$. If that was the case the value of ${\cal N}^{\rm ren}_{\infty}$ 
 would be analogous in some sense to the $g$-factor associated with 
1-dimensional RG defects of bulk flows\footnote{However while the $g$ factor of any interface is fixed by conformal symmetry alone 
that would not be the case for fixing the normalisation of a boundary condition changing operator.}.

Now that we have a precise construction of the local operator $\hat \psi_{0,\alpha}$ it is interesting to note that its OPE with itself 
is non-singular for finite $\alpha$. This stems from the fact that we can represent the norm squared of the perturbed  vacuum  as 
\be\label{norm}
{}_{\alpha}\langle 0 |0\rangle_{\alpha} = \lim_{\epsilon \to 0} \langle \hat \psi_{0,\alpha}(-\frac{\epsilon}{2}, 0) 
\hat \psi_{\alpha,0}(\frac{\epsilon}{2}, 0) \rangle 
\ee
that equals $1$ because the Bogolyubov transformation at hand is proper and $\hat \psi_{0,\alpha}(0,0)$ is a 
unitary operator. 
We illustrate (\ref{norm}) on the following figure. 

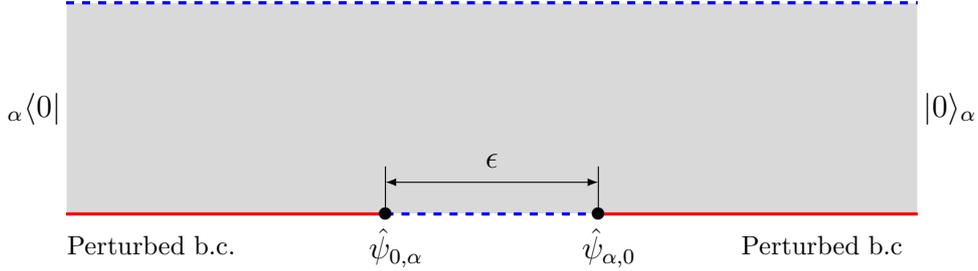
\begin{center}
\begin{tikzpicture}[>=latex,scale=1.4]
\filldraw[fill=gray!30!white,draw=white] (-4,0) rectangle (4,2);

\draw[blue,very thick,dashed](-1,0)--(1,0);
\draw[red,very thick] (-4,0)--(-1,0);
\draw[red,very thick] (1,0)--(4,0);
\draw[blue, very thick, dashed] (-4,2)--(4,2);
\draw (-1,0) node {$\bullet$} ;
\draw (1,0) node {$\bullet$} ;
\draw (4.3,1) node {$|0\rangle_{\alpha}$};
\draw (-4.3,1) node {${}_{\alpha}\langle 0|$};
\draw (1.1,-0.32) node {$\hat \psi_{\alpha, 0}$};
\draw (-0.9,-0.32) node {$\hat \psi_{0,\alpha}$};
\draw (-3.2,-0.3) node {{\small Perturbed b.c.}};
\draw (3.1, -0.3) node {{\small Perturbed b.c}} ;
\draw (-1,0) -- (-1, 0.45);
\draw (1,0)--(1,0.45);
\draw[<->] (-1, 0.3) -- (1,0.3);
\draw (0,0.5) node {$\epsilon$};
\end{tikzpicture}
\end{center}
Similarly we can swap the operators $\hat \psi_{\alpha, 0}$, $\hat \psi_{0,\alpha}$ and obtain a representation of 
the norm squared of $|0\rangle$ as represented in the perturbed theory Fock space. That norm squared also equals to 1. 

For $\alpha=\infty$ we had to renormalise the prefactor ${\cal N}_{\alpha}$ to obtain a primary operator whose OPE with itself 
is singular. 
This reflects the fact the Bogolyubov transformation at $\alpha=\infty$ is not proper. The limits $\alpha\to \infty$ and $\epsilon \to 0$ 
do not commute as simply reflects the general fact that the UV behaviour changes discontinuously at the IR fixed point.

It is also interesting to investigate the leading corrections to the asymptotic of $\hat \psi_{0,\alpha}$ in the limit 
$\alpha \to \infty$. Besides the normalisation factor ${\cal N}_{\alpha}$ the coupling $\alpha$ enters into (\ref{psi_hat1}) 
only via the quantities $d_{p}$ and $\tilde d_{p}$. Assuming $\alpha \gg p$ and $\alpha \gg k$ we can find the leading asymptotic using 
the asymptotic of the integral 
\be
 I(\alpha, \omega) = \int\limits_{0}^{\infty}\frac{dx}{x^2 + \omega^2}{\rm ln}\left( 1 + \frac{x}{2\pi \alpha^2 \tanh(\pi x)} \right) 
\sim \frac{{\rm \ln} \alpha}{\pi \alpha^2} \, .
\ee
This gives the leading asymptotics 
\be
d_{k} \sim g(k){\rm exp}\Bigl(  -(k+1/2)\frac{{\rm \ln} \alpha}{(\pi \alpha)^2} \Bigr) \, , \quad 
\tilde d_{p} \sim g(p){\rm exp}\Bigl(  p\frac{{\rm \ln} \alpha}{(\pi \alpha)^2} \Bigr) \, 
\ee
that means that up to a change in normalisation, the leading correction to $\hat \psi_{0,\infty}$ is 
 \be \label{hat_psi_asympt}
\hat \psi_{0,\alpha} \sim  {\rm exp}\Bigl(  -\frac{{\rm \ln} \alpha}{(\pi \alpha)^2}L_{0} \Bigr) \hat \psi_{0,\infty} 
{\rm exp}\Bigl(  \frac{{\rm \ln} \alpha}{(\pi \alpha)^2}L_{0} \Bigr) =  {\rm exp}\Bigl(  -\frac{{\rm \ln} \alpha}{(\pi h)^2}\partial_{\tau} \Bigr) \hat \psi_{0,\infty}
\ee
that is just a shift in the position of $\hat \psi_{0,\infty}$. The operator is shifted away from the ultraviolet boundary condition into the region of the 
infrared boundary condition. 
We can explain the appearance of logarithm in this expression using the infrared effective description of the theory near the $\alpha=\infty$ fixed point.  
It was shown in \cite{Toth} that near the infrared fixed point the theory is described in the leading order by a perturbation of the Euclidean 
action 
\be \label{Toth_effact}
S_{\alpha} = S_{\rm IR} + \frac{1}{2\pi h^2} \int\limits_{-\infty}^{\infty}\!\! d\tau\, [T_{ww}(\tau) + T_{\bar w \bar w}(\tau)] 
\ee
where $T_{ww}(\tau)$, $T_{\bar w\bar w}(\tau)$ stand for the stress energy tensor components restricted to the $\sigma=0$ boundary.
We note that (\ref{Toth_effact}) corresponds to the perturbation of dimensionless Hamiltonian (see (\ref{Hgen}), (\ref{theta}))
\be
h_{\alpha} = h_{\rm IR} - \frac{1}{2\pi^2\alpha^2}T(1)
\ee 
where $T=-2\pi T_{zz}$ stands for the stress-energy  component on the half-plane.  

 At the leading order in conformal perturbation 
theory the operator $\hat \psi_{0,\alpha}(0,0)$ can be described as the operator $\hat \psi_{0,\infty}$ with a single insertion of 
$$
-\frac{1}{\pi h^2} \int\limits_{-\infty}^{0}\!\! d\tau\, T_{ww}(\tau) \, .
$$
The contribution from short distances then comes by integrating the singular terms in the operator product expansion
\be
T_{ww}(\tau) \hat \psi_{0,\infty}(0) = -\frac{1}{2\pi}\Bigl[ 
\frac{ \hat \psi_{0,\infty}(0)}{16\tau^2} + \frac{\partial_{\tau}\hat \psi_{0,\infty}(0)}{\tau} + \dots \Bigr]\, .
\ee
Introducing a long distance cutoff $l$ and a short distance cutoff $\epsilon$ we obtain 
\be
- \frac{1}{\pi h^2} \int\limits_{-l}^{-\epsilon}\!\! d\tau\, T_{ww}(\tau)  \hat \psi_{0,\infty}(0) = 
  \frac{1}{2\pi^2 h^2}\left( \frac{ \hat \psi_{0,\infty}(0)}{16\epsilon} - {\rm ln}(\epsilon/l) \partial_{\tau}\hat \psi_{0,\infty}(0)\right)  + \dots 
\ee
where we explicitly write only the UV-divergent terms. 
The latter have to be subtracted by counterterms which define the renormalised operator $\hat \psi_{0, \alpha}$. 
The term that diverges as $1/\epsilon$ contributes to the overall normalisation of $\hat \psi_{0, \alpha}$ while the logarithmic divergence 
is cancelled by a counterterm 
\be
\hat \psi^{\rm c.t.} = -\frac{1}{2\pi^2 h^2}  {\rm ln}(\epsilon h^2 ) \partial_{\tau}\hat \psi_{0,\infty}(0) 
\ee
where $h^2$ is inserted inside the logarithm for dimensional reasons. 
Choosing $l=L/\pi$ (that amounts to just choosing a particular subtraction point) we obtain a leading order correction
\be
\hat \psi_{0, \alpha} =  \hat \psi_{0,\infty} -\frac{{\rm ln}(\alpha)}{\pi^2 h^2}\partial_{\tau}\hat \psi_{0,\infty}(0) 
\ee
that  
gives the same shift as in (\ref{hat_psi_asympt}) . 

Note that  the generator of  translations along the $\sigma$ direction  is given by
\be
H^{\perp}=  -\int\limits_{-\infty}^{\infty}[T_{ww}(\tau) + \bar T_{\bar w \bar w}(\tau)]d\tau \, .
\ee  
Its insertion on the boundary generates downward  shifts of the boundary (making the strip wider). 
The leading perturbation (\ref{Toth_effact})  then corresponds to shifting the $\sigma =0$ boundary downwards by $\frac{1}{4\pi^2 \alpha^2}$.
The signs of both shifts are important.   Both are such that singularities coming from collisions of bulk operators with the perturbed boundary and 
from boundary operators inserted in the unperturbed boundary colliding with the interface get smoothed out. The shift related to 
(\ref{Toth_effact}) has been used to construct regularised trial vacuum states in a variational method of \cite{Cardy_var}. 
Similarly the shift analogous to (\ref{hat_psi_asympt}) can be used in a variational method for boundary RG flows. The author 
is planning to elaborate on this idea in \cite{AK_inprogress}. 

\section{The mode truncated theory} \label{Modtr_sec}
\setcounter{equation}{0}

Here we consider a regulated version of  theory (\ref{halpha}) in which we restrict all mode sums to run from $k=0$ to $k=n_{c}$. 
We call this regularisation mode truncation. It was first considered in \cite{Toth}.
To define the mode-truncated Hilbert space ${\cal H}_{L}^{n_c}$ we first  define a finite-dimensional  Fock space representation ${\cal F}_{L}^{n_c}$
  for the canonical  oscillator algebra spanned by 
$\{A,A^{\dagger},a_{k+1/2},a^{\dagger}_{k+1/2}|k=0,1,\dots n_c\}$ with the Fock vacuum $|0\rangle$. The physical 
state space ${\cal H}_{L}^{n_c}$ is obtained by acting on the vacuum by products of even numbers of  creation operators. 
Its dimension is $2^{n_c+1}$.  

The mode truncated Hamiltonian reads 
\be \label{halphag1}
 h_{\alpha}^{n_c}   =  - \frac{1}{48}+ \sum_{k=0}^{n_c}(k+\frac{1}{2})a^{\dagger}_{k+1/2}a_{k+1/2}  + i\alpha a \sum_{k=0}^{n_c}(a^{\dagger}_{k+1/2} + a_{k+1/2}) \, .
\ee
 The spectral equation is
\be
\sum_{k=0}^{n_c}\frac{2\omega^2}{(k+1/2)^2 - \omega^2} = -\frac{\omega^2}{2\alpha^2} \, .
\ee
This equation has non-negative solutions 
\be 
\tilde \omega_{\alpha, 0}= 0<\tilde \omega_{\alpha,1}< ... < \tilde \omega_{\alpha, n_{c}+1}
\ee
 where 
$ i-1/2 <\tilde \omega_{\alpha,i}<i+1/2$, $i=1, \dots, n_c$ and $\tilde \omega_{\alpha, n_c+1}>n_{c}+1/2$. For $\alpha \gg \frac{(2n_c+ 1)}{4\sqrt{n_c+1}}$ we have 
\be\label{Omega_asympt}
\tilde \omega_{\alpha, n_c+1}\equiv \Omega_{\alpha} \to 2|\alpha|\sqrt{n_c+1} \, .
\ee
Thus we have a zero mode (as in the untruncated theory), a number of light modes approximating the first $n_c$ positive solutions $\omega_{n}$
 of the untruncated theory and a mode $\Omega_{\alpha}$ that becomes large for large $\alpha$. We can call the latter mode a heavy mode. 
 We denote the diagonalising modes respectively as $\tilde b_{\alpha,0}$, $\tilde b_{ \alpha,p}$, $p=1, \dots, n_c+1$ with 
 $\tilde b_{\alpha, n_c+1}\equiv b_{\alpha, \Omega}$ being the heavy mode. The corresponding Bogolyubov transformation is 
  \bea \label{Bog1tr}
\tilde b^{\dagger}_{\alpha, n} &=&\sum_{k=0}^{n_c}  (\tilde A_{n, k}a^{\dagger}_{k+1/2} + \tilde B_{n, k}a_{k+1/2}) + \frac{a}{\tilde f_{\alpha,n}}\, , \nonumber \\
\tilde b_{\alpha, n} &=& \sum_{k=0}^{n_c}  (\tilde A^{*}_{n, k}a_{k+1/2} + \tilde B^{*}_{n, k}a^{\dagger}_{k+1/2}) + \frac{a}{\tilde f_{\alpha,n}}
\eea
with 
\be 
 \tilde A_{n,k}= \frac{2i\alpha}{(k+1/2-\tilde \omega_{\alpha,n})\tilde f_{\alpha,n}} \, , \qquad 
\tilde B_{n,k}= \frac{-2i\alpha}{(k+1/2+\tilde \omega_{\alpha,n})\tilde f_{\alpha,n}}\, 
\ee
where  $\tilde f_{\alpha, n}>0$ and
\be\label{tildef}
 (\tilde f_{\alpha, n})^2   = 2+ 4\alpha^2 \sum_{k=0}^{n_c}\left( \frac{1}{(k+1/2-\tilde \omega_{\alpha,n})^2}  + 
  \frac{1}{(k+1/2+\tilde \omega_{\alpha,n} )^2}  \right) \, .
\ee
 The zero mode is chosen to be
\be\label{Bog2tr}
\tilde b_{\alpha, 0} =  \sum_{k=0}^{n_c} (\tilde A_{0, k}a^{\dagger}_{k+1/2} + \tilde B_{0, k}a_{k+1/2}) + \frac{a}{\tilde f_{\alpha,0}}
\ee
with 
\be
\tilde A_{0,k}=\frac{2i\alpha}{(k+1/2)\tilde f_{\alpha,0}}\, , \quad \tilde B_{0,k}= \frac{-2i\alpha}{(k+1/2)\tilde f_{\alpha,0}} 
\ee
and $\tilde f_{\alpha,0}>0$ given by (\ref{tildef}) with $\tilde \omega_{\alpha,n}$ replaced by zero. 

  
 For $\alpha\to \infty$ the light modes approach solutions to 
 \be\label{spec_inf}
 \sum_{k=0}^{n_c}\frac{1}{(k+1/2)^2 - \omega^2} = 0
 \ee
that are located near positive integers.  For $n\ll n_c$ we have   solutions to (\ref{spec_inf})
\be \label{IRspec_tr}
\tilde \omega_{n} = n + \frac{2n}{(n_c+3/2)\pi^2} + {\cal O}[\left(\frac{n}{n_{c}+3/2}\right)^2] \, . 
\ee
We denote the light modes at $\alpha=\infty$ as $\tilde b_{0}$, $\tilde b_{n}^{\dagger}$, $\tilde b_{n}$, $n=1, \dots, n_c$. 
From (\ref{IRspec_tr}) we see that  the boundary magnetic field theory mode-truncated at the UV fixed point  flows in the far infrared to a fixed spin boundary CFT regulated 
in some more complicated way.  The UV truncation parameter  $n_c$ controls  this regularisation of the IR theory 
in the sense that for $n_c\to \infty$ we recover the correct continuum spectrum.  In principle one can  find an effective IR Hamiltonian 
using some other regularisation scheme, say the mode truncation in the IR fermion modes, that reproduces (\ref{IRspec_tr}) and 
higher order corrections. 


We can generalise (\ref{halphag1})  by adding perturbations by other local operators bilinear in $\psi$ keeping the theory solvable 
by Bogolyubov transformation. 
We choose to add a perturbation by the stress energy tensor operator 
$T=-\frac{L}{2\pi} :\! \psi\partial_{\tau} \psi\!:(0,0)  $  with a coupling $g$. This gives us the following dimensionless Hamiltonian 
\bea \label{halphag}
&& h_{\alpha,g}^{n_c}   =  - \frac{1}{48}+ \sum_{k=0}^{n_c}(k+\frac{1}{2})a^{\dagger}_{k+1/2}a_{k+1/2}  + i\alpha a \sum_{k=0}^{n_c}(a^{\dagger}_{k+1/2} + a_{k+1/2}) \nonumber \\
&& - \frac{g}{2}:\!\left(\sum_{k=0}^{n_c}\sum_{l=0}^{n_c}( a^{\dagger}_{k+1/2} +a_{k+1/2})l(a^{\dagger}_{l+1/2} - a_{l+1/2})\right)\!: \, .
\eea
We can find the spectrum of $h_{\alpha,g}^{n_c}$ by writing an ansatz of the same form as (\ref{B1}) and then substituting it into 
(\ref{com_eq}). This gives a system of linear equations that has a nontrivial solution provided that $\omega$ solves the following transcendental 
equation 
\be \label{alphag_spec}
\sum_{k=0}^{n_c}\frac{2\omega^2}{(k+1/2)^2 - \omega^2} = -\frac{\omega^2(1+g(n_c+1))^2}{\omega^2 g(1+\frac{g}{2}(n_c+1)) +2\alpha^2}  \, .
\ee
Solutions to this spectral equation give energy gaps corresponding to creation operators diagonalising $h_{\alpha,g}^{n_c} $.

We will be also interested in the theory obtained by perturbing the infrared fixed point by the stress-energy tensor. In this case 
we are dealing with Ramond free fermions with Hamiltonian (\ref{h_inf}). The mode truncated theory perturbed by $T(0,0)$ is described by 
the Hamiltonian 
\be\label{g_Ham2}
\tilde h^{n_c}_{g} = \frac{1}{24} + \sum_{k=1}^{n_{c}}k\, b^{\dagger}_{k}b_{k} 
- \frac{g}{2}:\!\left(\left[ \sum_{k=1}^{n_c} (b^{\dagger}_{k} +b_{k}) + b_{0}\right] \sum_{l=1}^{n_c} l(b^{\dagger}_{l} - b_{l})\right)\!: \, .
\ee
It can be diagonalised by a Bogolyubov transformation that gives a spectral equation 
\be\label{g_spec2}
\sum_{k=1}^{n_c}\frac{2\omega^2}{k^2 - \omega^2} -1=-\frac{(1+\frac{g}{2}(1+2n_c))^2}{g(1+g(1+2n_c)/4)} \, .
\ee

The spectra of both theories perturbed by the stress energy tensor demonstrate some peculiar features. 
Looking for example at (\ref{g_spec2}) we see that for   $g\ll 1/n_c\ll 1$   the low lying spectrum is well approximated 
by the equation 
\be
\tan(\pi \omega) = g\pi \omega 
\ee 
In particular we see that choosing $g=-\frac{1}{2\pi^2 \alpha^2}$ we reproduce (\ref{spec}). Taking a small negative $g$ and  increasing 
it in magnitude to the value  $g_{*}=-\frac{2}{1+2n_c}$ generates a reversed RG flow in which we approach approximately half-integer 
values that is the spectrum of the UV fixed point. Increasing the magnitude further will shift the spectrum back towards the integer 
values which we reach again at $g^{*}=-\frac{4}{1+2n_c}$. This looks like a bounce flow akin to the ones observed in TCSA numerics  
for boundary RG flows \cite{Feverati_etal2}, \cite{Toth}, \cite{Gerard2}. Further increase of $g$ towards minus infinity does not change the low lying spectrum 
much at all. If we choose to take positive $g$ then qualitatively the behaviour is monotonic and we reach the same end point as going 
towards minus infinity. It should be noted that while the solutions to the spectral equation are always real for real 
coupling $g$ for positive $g$ and for  large negative values of $g$ the spectrum looks like (\ref{spec}) with imaginary values of $\alpha$. 
We suspect that this signifies a certain instability of the vacuum similar to the one observed in the $T\bar T$-perturbed bulk 
theories \cite{Zam_TT}, \cite{Tateo_etal}. We are going to discuss these issues in more detail in \cite{KM2}.



\section{Effective Hamiltonian near the UV fixed point} \label{UV_sec}
\setcounter{equation}{0}

The spectrum of the mode truncated theory differs from that of the continuum theory with corrections suppressed by inverse powers of $n_c$. 
We would like to construct an effective Hamiltonian in the truncated Fock space that improves the accuracy of approximation 
of the spectrum for the low lying energies. Such an effective Hamiltonian can be expanded in (truncated) irrelevant operators 
with coefficients containing positive powers of $\alpha$ and negative powers of $n_c$. We will first describe a construction for 
such an effective Hamiltonian that uses the exact solution to theory (\ref{halphag}) and then will explain how such Hamiltonians 
can be systematically obtained by Schrieffer-Wolff method. 

To proceed with the first method we note that  the continuum frequencies $\omega_{n}$, that are solutions to the full untruncated spectral equation (\ref{spec}), can be represented as 
 solutions to a deformed truncated spectral equation by rewriting (\ref{spec}) as  
 \be \label{spec_def}
 \sum_{k=0}^{n_c}\frac{2\omega^2}{(k+1/2)^2 - \omega^2} = -\frac{\omega^2}{2\alpha^2} -  
 \sum_{k=n_c+1}^{\infty}\frac{2\omega^2}{(k+1/2)^2 - \omega^2} \, .
 \ee
 For $\omega< n_c+3/2$ we can obtain approximations to the exact frequencies by expanding the right hand side in powers of $\omega/(n_c+3/2)$. 
 The extra terms on the right hand side can be matched to couplings of irrelevant operators in the effective Hamiltonian. 
 Thus expanding the right hand side of  (\ref{spec_def}) to the  order $\omega^4$ we obtain 
 \be\label{spec_expand}
  \sum_{k=0}^{n_c}\frac{2\omega^2}{(k+1/2)^2 - \omega^2} = -\frac{\omega^2}{2\alpha^2} - 2\omega^2{\cal C}_{n_c}^{(2)} -2\omega^4  {\cal C}_{n_c}^{(4)}
  + \dots 
 \ee
 where 
 \be\label{Cnc}  
{\cal C}_{n_c}^{(4+2p)} = \sum_{l=n_c+1}^{\infty}\frac{1}{(l+1/2)^{4+2p}} = \frac{1}{(3+2p)(n_c+3/2)^{3+2p}}  + {\cal O} \left(\frac{1}{(n_c+3/2)^{4+2p}} \right).
\ee
Comparing (\ref{spec_expand}) with the expansion to the order $\omega^4$ of the right hand side of (\ref{alphag_spec}) with the couplings 
 $\alpha^{\rm eft}$, $g^{\rm eft}\equiv g(\alpha)$ we find matching when  
 \bea
&&  \frac{(1+g^{\rm eff}(n_c+1))^2}{2(\alpha^{\rm eff})^2} = \frac{1}{2\alpha^2} + 2{\cal C}_{n_c}^{(2)}\, , \nonumber \\
&& \frac{g(1+g(n_c+1)/2))(1+g(n_{c}+1))^2}{4(\alpha^{\rm eff})^4} = -2 {\cal C}_{n_c}^{(4)}  \, .
 \eea
Solving these equations we obtain  
 \be
 g^{\rm eff}\equiv g(\alpha)=\frac{1}{n_c+1}\left( \frac{1}{\sqrt{1+{\cal K}}}-1 \right)\, , \quad \mbox{where } 
 {\cal K} = -\frac{8\alpha^{4} {\cal C}_{n_c}^{(4)}}{(1+4\alpha^2 {\cal C}_{n_c}^{(2)})^2}
 \ee
 and 
 \be
\alpha^{\rm eff} =  \frac{\alpha[1+g(\alpha)(n_c+1)]}{\sqrt{1+4{\cal C}_{n_c}^{(2)}\alpha^2}}  
 \ee
 each defined up to order $1/(n_{c})^4$ corrections. Expanding in powers of $\alpha$ we obtain 
  \be \label{galpha_exp2}
 g(\alpha)  = -8 {\cal C}_{n_c}^{(4)}\alpha^4  +  {\cal O}(\alpha^6/n_c^4) \, , \quad \alpha^{\rm eff} = \alpha -2{\cal C}_{n_c}^{(2)}\alpha^3 + {\cal O}(\alpha^5/n_c^2)\, .
 \ee
 For this solution we observe that terms at higher powers of $\omega^2$ in (\ref{alphag_spec}) are suppressed by at least $1/(n_{c})^6$. 
  This means that the theory (\ref{halphag}) with $\alpha=\alpha^{\rm eff}$ and $g=g(\alpha)$ as above has the low lying spectra $\omega_{n}\ll n_c$ 
 coinciding with the continuous theory spectra defined at $\alpha < n_{c}+3/2$ up to terms of  order $1/(n_c)^5$. 
 Thus in fact one can expand the effective couplings to the order $1/(n_{c})^4$ and truncate the expansion. 
 This derivation of the effective Hamiltonian is somewhat ad hoc, it relied on knowing the exact spectrum for theory (\ref{halphag}).
 To improve the approximation we need to include higher dimension operators into the effective Hamiltonian.
 Next  we are going derive the same effective Hamiltonian 
 through the order $\alpha^4$ in the coupling using the Schrieffer-Wolff method that is a systematic method for producing 
 approximations to any order of accuracy.

We can represent  the Hamiltonian of the continuous theory as 
\be
h_{\alpha} =  h_{0}^{\rm L} + h_{0}^{\rm H} + v^{\rm L} + v^{\rm LH}
\ee
where 
\be
h_{0}^{\rm L}= \sum_{k=0}^{n_c}(k+\frac{1}{2})a^{\dagger}_{k+1/2}a_{k+1/2} - \frac{1}{48}\, , \qquad 
h_{0}^{\rm H}= \sum_{k=n_c + 1}^{\infty}(k+\frac{1}{2})a^{\dagger}_{k+1/2}a_{k+1/2} \, , 
\ee
\be
v^{\rm L}= i\alpha a \sum_{k=0}^{n_c}(a^{\dagger}_{k+1/2} + a_{k+1/2})\, ,  \qquad 
 v^{\rm LH}=i\alpha a \sum_{k=n_c+1}^{\infty}(a^{\dagger}_{k+1/2} + a_{k+1/2})
\ee
so that the index L labels contributions of light modes with $k=0,\dots , n_c$ and H labels those of the heavy modes with $k>n_c$. 
The term $v^{\rm LH}$ mixes the light and heavy modes. 

The general idea of Schrieffer-Wolff method \cite{SW} is to find a sequence of unitary rotations in ${\cal H}_{L}$ that replaces the off-diagonal 
terms that mix the light and heavy modes by new off-diagonal terms suppressed by additional negative powers of $n_c$ and positive powers of $\alpha$ 
at the expense of new interactions between the light modes.

At the first step we remove $v^{\rm LH}$ by a unitary transformation 
\be
h_{\alpha} \mapsto h_{\alpha, 1} = U_1 h_{\alpha} U_{1}^{-1}
\ee
where $U_{1}=e^{T_1}$ is chosen so that 
\be \label{tr_choice}
[T_1,h_{0}^{\rm H}]= - v^{\rm LH} \, .
\ee
This can be achieved by taking 
\be\label{integration}
T_{1} = \int\limits_{-\infty}^{0}\! dt\, e^{t(h_{0}^{\rm L} + h_{0}^{\rm H})} v^{\rm LH}  e^{-t(h_{0}^{\rm L} + h_{0}^{\rm H})} 
\ee
that for the case at hand gives
\be\label{T1}
T_1= i\alpha a \sum_{k=n_c+1}^{\infty}\frac{a^{\dagger}_{k+1/2} - a_{k+1/2}}{k+1/2}\, . 
\ee
Note that $T_1 \sim \alpha/n_c$. Thus the new interaction term in $h_{\alpha, 1}$ that mixes the light and heavy modes will be at least 
of order $(\alpha/n_{c})^2$. We can remove it by another unitary transformation constructed as in (\ref{tr_choice}), (\ref{integration}), that 
will push the mixing interaction to a yet higher order. Repeating this procedure we can obtain (assuming convergence of the method) 
a sequence of effective Hamiltonians with better and better approximations of the continuum spectrum.

For the transformation generated by (\ref{T1}) we calculate through the order $\alpha^4$
\be
 h_{\alpha,1} = {\cal E}_{1}(\alpha, n_c)  + h_{0}^{\rm L} + h_{0}^{\rm H} + v^{\rm L} _1 + v^{\rm H}_{1} + v^{\rm LH}_{1} + {\cal O}(\alpha^5)
\ee
where ${\cal E}_{1}(\alpha, n_c)$ is the shifted vacuum energy in which we have little interest, and   
\be
v^{\rm L}_1 = i(\alpha - 2\alpha^3 {\cal  C}_{n_c}^{(2)} ) a \sum_{k=0}^{n_c}(a^{\dagger}_{k+1/2} + a_{k+1/2})\, , 
\ee
\be
v^{\rm H}_{1}= (-\alpha^2 + \alpha^4 {\cal C}_{n_c}^{(2)})  
\sum_{l=n_c+1}^{\infty}(a^{\dagger}_{l+1/2} + a_{l+1/2})  \sum_{k=n_c+1}^{\infty} \frac{a^{\dagger}_{k+1/2} - a^{\dagger}_{k+1/2}}{k+1/2}  \, , 
\ee
\bea
&& v^{\rm LH}_{1} = (-2\alpha^2 + \frac{4}{3}{\cal C}_{n_c}^{(2)}\alpha^4) \sum_{l=0}^{n_c}(a^{\dagger}_{l+1/2} + a_{l+1/2})  \sum_{k=n_c+1}^{\infty} \frac{a^{\dagger}_{k+1/2} - a^{\dagger}_{k+1/2}}{k+1/2}  \nonumber \\
&& -i\frac{4}{3}\alpha^3 {\cal C}_{n_c}^{(2)} a \sum_{l=n_c+1}^{\infty}(a^{\dagger}_{l+1/2} + a_{l+1/2}) 
\eea
where ${\cal C}_{n_c}^{(2)} $ is given in (\ref{Cnc}).
In the new Hamiltonian the mixing interaction $v_{1}^{\rm LH}$ starts now with the order $\alpha^2$. We push this to the next order 
by performing another  unitary transformation 
\be
h_{\alpha,1} \mapsto h_{\alpha, 2} = U_2 h_{\alpha,1} U_{2}^{-1} \, , \qquad  U_{2}=e^{T_2} 
\ee
where $T_2$ is chosen so that it  removes the term of the order $\alpha^2$ in $v_{1}^{LH}$, i.e. 
\be 
[T_2,h_{0}^{\rm H}]= 2\alpha^2 \sum_{l=0}^{n_c}(a^{\dagger}_{l+1/2} + a_{l+1/2})  \sum_{k=n_c+1}^{\infty} \frac{a^{\dagger}_{k+1/2} - a^{\dagger}_{k+1/2}}{k+1/2}  \, .
\ee
Using the construction (\ref{tr_choice}), (\ref{integration}) again we find 
\be
 T_2= -2\alpha^2  \sum_{k=0}^{n_c} \sum_{l=n_c+1}^{\infty} \Bigl( 
\frac{a^{\dagger}_{k+1/2}a^{\dagger}_{l+1/2} + a_{k+1/2}a_{l+1/2}}{(l+1/2)(k+l+1)} 
+ \frac{a_{k+1/2}a^{\dagger}_{l+1/2} + a_{k+1/2}^{\dagger}a_{l+1/2}}{(l+1/2)(l-k)}      \Bigr) \, . 
\ee
 For the new effective Hamiltonian we obtain 
\be
 h_{\alpha,2} = {\cal E}_{2}(\alpha, n_c)  + h_{0}^{\rm L} + h_{0}^{\rm H} + v^{\rm L} _2 + v^{\rm H}_{2} + v^{\rm LH}_{2} + {\cal O}(\alpha^5)
\ee
where 
\bea \label{eff_light}
&&v^{\rm L}_2 = i(\alpha - 2\alpha^3 {\cal  C}_{n_c}^{(2)} ) a \sum_{k=0}^{n_c}(a^{\dagger}_{k+1/2} + a_{k+1/2}) 
\nonumber \\
&& + 2\alpha^4 \sum_{n=0}^{n_c}(a^{\dagger}_{n+1/2} + a_{n+1/2})\sum_{k=0}^{n_c} (a^{\dagger}_{k+1/2} - a_{k+1/2}) f_{n_c}(k) \, , 
\eea
\be
f_{n_c}(k) = \sum_{l=n_c+1}^{\infty} \frac{2k+1}{(l-k)(k+l+1)(l+1/2)^2} \, .
\ee
The function $f_{n_c}(k)$ can be expanded in powers of $(k+1/2)$. The expansion contains only odd powers and 
the coefficient at $(k+1/2)^p$ is of order $(1/n_{c})^{p+2}$.
At the leading order one has 
\be 
f_{n_c}(k) = 2{\cal C}_{n_c}^{(4)}(k+1/2) + {\cal O}((k+1/2)^3) \, .
\ee

The terms in $v^{\rm L}$ now start from the order $\alpha^3$. We can push this to the order ${\cal O}(\alpha^5)$ by 
applying two more unitary transformations $e^{T_{3}}$ and $e^{T_{4}}$ with $T_3 \sim \alpha^3$ and $T_4\sim \alpha^4$. 
Direct inspection shows that these additional transformations do not induce any terms of order less than $\alpha^5$ in the 
light modes interaction $v^{\rm L}$.  Hence through the order $\alpha^4$ the effective Hamiltonian for the light modes 
is given in (\ref{eff_light}). It contains the original magnetic field interaction with an effective coupling 
\be\label{effalpha2}
\alpha^{\rm eff} = \alpha - 2\alpha^3 {\cal  C}_{n_c}^{(2)}  
\ee 
as well as an infinite sequence of couplings to  truncated irrelevant operators 
\be\label{Otr_def}
{\cal O}_{p}^{\, \rm tr} = -\frac{1}{2} \left(\frac{L}{\pi}\right)^{p+1} :\! \psi \partial_{\tau}^{p}\psi \!:(0,0)
  -\frac{1}{2}\sum_{n=0}^{n_c}:\! (a^{\dagger}_{n+1/2} + a_{n+1/2})\sum_{k=0}^{n_c}  k^p (a^{\dagger}_{k+1/2} - a_{k+1/2}) \!: 
\ee
where $p$ is odd  and  the leading operator ${\cal O}_{1}^{\, \rm tr}$ is the truncated holomorphic stress-energy tensor $T^{\, \rm tr}$ 
component taken on the half-plane .  

If we continue the Schrieffer-Wolff process  it is not hard to see that the effective Hamiltonian is always of the form 
\bea \label{genUVeff}
&& h^{\rm eff}_{\alpha}  =  {\cal E}(\alpha, n_c) + \sum_{k=0}^{n_c}(k+\frac{1}{2})a^{\dagger}_{k+1/2}a_{k+1/2}  +
 i\alpha^{\rm eff}(\alpha, n_c) a \sum_{k=0}^{n_c}(a^{\dagger}_{k+1/2} + a_{k+1/2}) \nonumber \\
&& + g^{\rm eff}(\alpha, n_c) T^{\, \rm tr}(0,0) + \sum_{p=1}^{\infty} g^{\rm eff}_{2p+1}(\alpha, n_c) {\cal O}_{2p+1}^{\, \rm tr}(0,0) 
\eea
where each effective coupling has a double expansion in positive powers of $\alpha$ and negative powers of $n_{c}$. 
In our calculation that lead to (\ref{eff_light}) we found all terms in these expansions  through the order $\alpha^4$ each to all orders in $n_c$.  
These are given by  (\ref{effalpha2}) and  
\be
g^{\rm eff} =  -8 {\cal C}_{n_c}^{(4)}\alpha^4  + {\cal O}(\alpha^6) \, , \quad g^{\rm eff}_{2p+1} = -8 \alpha^4 {\cal C}_{n_c}^{(4+2p)} \, , \enspace p\ge 1
+ {\cal O}(\alpha^6) \, . 
\ee
The  explicit terms in $\alpha^{\rm eff}$ and $g^{\rm eff}$ coincide with those written in (\ref{galpha_exp2}) that were obtained by a different method. 

The effective Hamiltonian (\ref{genUVeff}) reproduces the spectrum of the continuous theory in the truncated theory. By moving all interaction terms 
into the left hand side of (\ref{genUVeff}) we obtain a representation of the truncated theory as the continuous theory plus corrections. This representation looks akin to 
the effective  Hamiltonian discussed in \cite{Lukyanov}, \cite{LTerras} that reproduces lattice size corrections within a continuum theory.

\section{Effective Hamiltonian near the IR fixed point}\label{IR_sec}
\setcounter{equation}{0}

We would like to discuss next how to construct an effective Hamiltonian on the RG trajectory corresponding to the 
boundary magnetic field model  near the infrared fixed point. In the language of effective action such QFTs are 
described by adding irrelevant interactions to the infrared CFT action. This is also the case for effective Hamiltonians. 
For the theory at hand the approach to IR was investigated in \cite{Toth} who showed that the leading interaction 
in the effective IR Hamiltonian is given by the stress energy tensor with a coupling 
\be
g^{\rm eff}_{\rm IR} = -\frac{1}{2\pi^2\alpha^2} \, .
\ee 
Subleading terms were not discussed in that paper. We are going to apply the Schrieffer-Wolff method using the regularisation scheme 
that comes from the mode truncation  
at the UV fixed point. This will allow us to derive subleading terms systematically and we will compute some of them explicitly.

As a first step we would like to express $h_{\alpha, n_c}$ in terms of the creation and annihilation operators $\tilde b_{n}^{\dagger}$, $\tilde b_{n}$ 
of the infrared CFT using Bogolyubov transformation (\ref{Bog1tr}), (\ref{Bog2tr}). To this end we first note that 
\be
h_{\beta, n_{c}} = h_{0}^{L} +  i\beta a \sum_{k=0}^{n_c}(a^{\dagger}_{k+1/2} + a_{k+1/2}) = \sum_{k=1}^{n_c}\tilde \omega_{\beta, k} \tilde b_{\beta, k}^{\dagger}\tilde b_{\beta, k} + \Omega_{\beta}
b^{\dagger}_{\beta, \Omega}b_{ \beta, \Omega} + {\cal E}_{\beta}(n_c) 
\ee
and thus we can write 
\be \label{h_rewrite}
h_{\alpha, n_c} = \sum_{k=1}^{n_c}\tilde \omega_{\beta, k} \tilde b_{\beta, k}^{\dagger}\tilde b_{\beta, k} + \Omega_{\beta}
b^{\dagger}_{\beta, \Omega}b_{ \beta, \Omega} + {\cal E}_{\beta}(n_c) +  i(\alpha-\beta) a \sum_{k=0}^{n_c}(a^{\dagger}_{k+1/2} + a_{k+1/2}) \, . 
\ee
We next rewrite the last term in (\ref{h_rewrite}) in terms of $\tilde b_{\beta,n}^{\dagger}$, $\tilde b_{\beta, n}$  using the inverse Bogolyubov 
transformation for (\ref{Bog1tr}), (\ref{Bog2tr}). This transformation is given by expressions of the same form as in (\ref{inverse1}),
 (\ref{inverse2}) with the Bogolyubov coefficients replaced by those given in (\ref{Bog1tr}), (\ref{Bog2tr}). In particular 
 we find using the inverse transformation that
\bea
&& a = 2\left( \frac{\tilde b_{\beta, 0}}{\tilde f_{\beta,0}}+ \sum_{k=1}^{n_c+1}\frac{\tilde b_{\beta, k}^{\dagger} +  \tilde b_{\beta, k} }{\tilde f_{\beta,k}}  \right) \, ,  
\nonumber \\
&& \sum_{k=0}^{n_c}(a^{\dagger}_{k+1/2} + a_{k+1/2}) = \frac{i}{\beta} \left(\sum_{k=1}^{n_c+1} 
\frac{\tilde \omega_{\beta, k}(\tilde b^{\dagger}_{\beta, k}  - \tilde b_{\beta, k}) }{\tilde f_{\beta,k}} \right)\, .
\eea
Substituting these expressions  into (\ref{h_rewrite}) we obtain 
\bea \label{halphabeta}
&& h_{\alpha, n_c}=\sum_{k=1}^{n_c}\tilde \omega_{\beta, k} \tilde b_{\beta, k}^{\dagger}\tilde b_{\beta, k} + \Omega_{\beta}
b^{\dagger}_{\beta, \Omega}b_{ \beta, \Omega} + {\cal E}_{\beta}(n_c) \nonumber \\
&& + 2\left(1-\frac{\alpha}{\beta}\right) \left(   \frac{\tilde b_{\beta, 0}}{\tilde f_{\beta,0}} + \sum_{k=1}^{n_c+1}\frac{\tilde b_{\beta, k}^{\dagger} +  \tilde b_{\beta, k} }{\tilde f_{\beta,k}} \right)\left( \sum_{l=1}^{n_c+1} 
\frac{\tilde \omega_{\beta, l}(\tilde b^{\dagger}_{\beta, l}  - \tilde b_{\beta, l}) }{\tilde f_{\beta,l}}   \right) \, .
\eea
In this expression we now take the limit $\beta \to \infty$. We find that the linearly divergent term $\Omega_{\beta}
b^{\dagger}_{\beta, \Omega}b_{ \beta, \Omega} $ cancels with a linearly divergent term from the last term in (\ref{halphabeta}). 
Also the linear divergent term in the vacuum energy is cancelled.
Noting the asymptotics 
\be
\tilde f_{\beta, n} \sim {\cal O}(\beta)\, , \enspace n=0, \dots, n_c \, \, , \quad \tilde f_{\beta, n_c+1} \to 2  
\ee
as $\beta \to \infty$, we obtain 
\be \label{halphanc}
h_{\alpha, n_c} = \tilde {\cal E}_{0}(n_c) + e_{0}^{\rm L} + e_{0}^{\rm H} + u^{\rm LH} 
\ee
with 
\be
\tilde {\cal E}_{0}(n_c) = -\frac{1}{48} + \frac{1}{2}\sum_{n=1}^{n_c}(n-\frac{1}{2} - \tilde \omega_{n}) + \frac{1}{2}(n_c + \frac{1}{2})\, , 
\ee
\be
e_{0}^{\rm L} = \sum_{k=1}^{n_c}\tilde \omega_{ k} \tilde b_{ k}^{\dagger}\tilde b_{ k} \, , \quad 
e_{0}^{\rm H} = 2 \sqrt{n_c+1}\alpha\,  {\rm sign}(\beta) \,  b^{\dagger}_{\Omega} b_{\Omega} \, , 
\ee
\be
u^{\rm LH}  = \sqrt{n_c+1}(b_{\Omega}-b^{\dagger}_{\Omega})
\left(\frac{\tilde b_{\beta, 0}}{t_{0}} + \sum_{k=1}^{n_c}\frac{\tilde b_{ k}^{\dagger} +  \tilde b_{ k} }{t_{k}}   \right)  \, . 
\ee
where 
\be
(t_{n} )^2= \sum_{k=0}^{n_c}\left( \frac{1}{(k+1/2-\tilde \omega_{n})^2}  +  \frac{1}{(k+1/2+\tilde \omega_{n} )^2}  \right) \, , \enspace t_{n}>0
\ee
and $n=0, \dots, n_c$.

It was not important for the above calculation but from now on we will assume that $\alpha$ has the same sign as $\beta$ as we clearly should in the context of the problem we are discussing. 
 With this assumption $\alpha {\rm sign}(\beta)=|\alpha|$ and the heavy mode $b^{\dagger}_{\Omega}$ has a positive excitation energy. 
Now we are in a very good position to apply the Schrieffer-Wolff method as we have a single heavy mode whose energy  grows 
linearly with $\alpha$ and creates a large energy gap between all light and heavy energy states. This is better than in the UV effective Hamiltonian 
case discussed in the previous section where the gap between the heaviest of the light states and the lightest of the heavy states was just 1.
It should be also noted that while the UV effective Hamiltonian (retaining all terms) reproduces the exact low energy spectrum of the 
continuum theory, our effective Hamiltonian near the infrared fixed point only aims at reproducing the truncated spectrum of $h_{\alpha, n_c}$. 

Before we present explicit calculations of the effective Hamiltonians it is worth organising the general expansion. Since our starting Hamiltonian 
is bilinear in the oscillator modes it is easy to see from (\ref{tr_choice}), (\ref{integration}) that it remains bilinear to all orders. By hermiticity a general term in
the effective Hamiltonian is a linear combination of a constant and  the following elementary bilinears
\be
N\equiv b^{\dagger}_{\Omega}b_{\Omega} \, , \quad (\tilde b_{n} +\tilde b_{n}^{\dagger})(\tilde b_{m} - \tilde b_{m}^{\dagger})\, , \quad 
\tilde b_{0}(\tilde b_{m} - \tilde b_{m}^{\dagger})
\ee
and 
\be
 B_{+}(\tilde b_{m} - \tilde b_{m}^{\dagger})\, , \quad B_{-}(\tilde b_{m} + \tilde b_{m}^{\dagger})\, , \quad 
 B_{-}\tilde b_{0}
\ee
where 
\be
B_{+}=b_{\Omega} + b^{\dagger}_{\Omega} \, , \quad B_{-}=b_{\Omega}^{\dagger}- b_{\Omega} \, .
\ee
A generator of a unitary transformation being antihermitean and containing only mixing terms  is a linear combination of 
\be
B_{-}(\tilde b_{m} - \tilde b_{m}^{\dagger})\, , \quad B_{+}(\tilde b_{m} + \tilde b_{m}^{\dagger})\, , \quad 
 B_{+}\tilde b_{0} \, . 
\ee
All commutators between two types of binomials are listed in Appendix B. Also when calculating generators $S$
satisfying 
\be
[S, e_{0}^{\rm L} + e_{0}^{\rm H}] = - u
\ee
we use 
\be
S = \int\limits_{-\infty}^{0}\!\! dt\, e^{t e_{0}} u  e^{-te_{0}} 
\ee
with $e_{0}=e_{0}^{\rm L} + e_{0}^{\rm H}$. Such expressions can be easily calculated by using 
\be\label{help1}
\int\limits_{-\infty}^{0}\!\! dt\, e^{t e_{0}} B_{+}(\tilde b_{m} - \tilde b_{m}^{\dagger}) e^{-te_{0}} =
\frac{W_{\alpha}B_{-}(\tilde b_{m} - \tilde b_{m}^{\dagger})
}{W_{\alpha}^2-\tilde \omega_{m}^2} 
+ \frac{\tilde \omega_{n}B_{+}(\tilde b_{m} + \tilde b_{m}^{\dagger})}{W_{\alpha}^2-\tilde \omega_{m}^2} \, , 
\ee
\be\label{help2}
\int\limits_{-\infty}^{0}\!\! dt\, e^{t e_{0}} B_{-}(\tilde b_{m} + \tilde b_{m}^{\dagger}) e^{-te_{0}} =
\frac{W_{\alpha}B_{+}(\tilde b_{m} + \tilde b_{m}^{\dagger})
}{W_{\alpha}^2-\tilde \omega_{m}^2} 
+ \frac{\tilde \omega_{n}B_{-}(\tilde b_{m} - \tilde b_{m}^{\dagger})}{W_{\alpha}^2-\tilde \omega_{m}^2} 
\ee
where 
\be
W_{\alpha} = 2|\alpha|\sqrt{n_c+1} \, .
\ee

The first unitary transformation we perform  on $h_{\alpha, n_c}$ is 
\be
h_{\alpha, n_c} \mapsto h_{\alpha, n_c}^{(1)} = e^{S_1}h_{\alpha, n_c}e^{-S_1}
\ee
with $S_1$ chosen so that 
\be \label{S11}
[S_{1}, e_{0}^{\rm L} + e_{0}^{\rm H}] = - u^{\rm LH} \, . 
\ee
Using (\ref{help1}), (\ref{help2}) we obtain 
\be
S_1 =  -\sqrt{n_c+1}\Bigl[   \sum_{k=1}^{n_c}\frac{1}{t_{k}}\left( \frac{W_{\alpha}B_{+}(\tilde b_{k} + \tilde b_{k}^{\dagger})}{W_{\alpha}^2 - \tilde \omega_{k}^2}
+  \frac{\tilde \omega_{k} B_{-}(\tilde b_{k} - \tilde b_{k}^{\dagger})}{W_{\alpha}^2 - \tilde \omega_{k}^2} \right) 
+ \frac{B_{+}\tilde b_{0}}{t_{0}W_{\alpha}}\Bigr]
\ee
Using (\ref{halphanc}), (\ref{S11}) we have 
\bea\label{expp}
&& h_{\alpha, n_c}^{(1)} = e^{S_1}( e_{0}^{\rm L} + e_{0}^{\rm H} -[S_{1}, e_{0}^{\rm L} + e_{0}^{\rm H}] )e^{-S_1} 
=  e_{0}^{\rm L} + e_{0}^{\rm H} - \sum_{n=1}^{\infty} \frac{n}{(n+1)!} (\hat S_{1})^{n+1} (e_{0}^{\rm L} + e_{0}^{\rm H}) \nonumber \\
&& = e_{0}^{\rm L} + e_{0}^{\rm H} -\Bigl[\frac{1}{2}(\hat S_{1})^2 + \frac{1}{3}(\hat S_{1})^3 + \frac{1}{8}(\hat S_{1})^4 + \frac{1}{30}(\hat S_{1})^5 
+ \dots \Bigr](e_{0}^{\rm L} + e_{0}^{\rm H})
\eea
where $\hat S_{1} X = [S_{1},X]$ stands for the adjoint action. Noting that $S_1\sim \frac{1}{\alpha}$ and $e_{0}^{\rm H}\sim \alpha$ 
we see that the term $\hat S^{n}(e_{0}^{\rm L} + e_{0}^{\rm H})$ is of order $1/(\alpha)^{n-1}$. To obtain all terms to the order $1/\alpha^4$ 
it suffices to keep terms in the expansion (\ref{expp}) up to $n=5$.
We obtain 
\be
h_{\alpha, n_c}^{(1)} = \tilde {\cal E}_{1}(\alpha, n_c) + e_{0}^{\rm L} + e_{1}^{\rm H} + u_{1}^{\rm L}  
+ u_{1}^{\rm LH}  + {\cal O}\left(\frac{1}{\alpha^5}\right)
\ee
\be
 \tilde {\cal E}_{1}(\alpha, n_c) = \tilde {\cal E}_{0}(n_c) -\frac{1}{2} (n_c+1){\cal D}_{\alpha}^{(1)}
 +\frac{1}{2}(n_c+1)^2{\cal D}_{\alpha}^{(1)} \left( {\cal D}_{\alpha}^{(2)} + \frac{{\cal D}_{\alpha}^{(1)} }{2W_{\alpha}} \right)
 \, , 
\ee
\be
e_{1}^{\rm H} = \Bigl[W_{\alpha} +(n_c+1){\cal D}_{\alpha}^{(1)}  -(n_c+1)^2{\cal D}_{\alpha}^{(1)} \left( {\cal D}_{\alpha}^{(2)} + \frac{{\cal D}_{\alpha}^{(1)} }{2W_{\alpha}} \right) \Bigr] b^{\dagger}_{\Omega}b_{\Omega} \, , 
\ee
\bea\label{u1L}
&& u_{1}^{\rm L} = (n_c+1) \left( \sum_{l=1}^{n_c}\frac{\tilde b_{l}+\tilde b_{l}^{\dagger}}{t_{l}} + \frac{\tilde b_{0}}{t_{0}} \right)
\left( \sum_{k=1}^{n_c}\frac{\tilde \omega_{k}(\tilde b_{k}^{\dagger} - \tilde b_{k})}{t_{k}(W_{\alpha}^2-\tilde \omega_{k}^2)}\right) 
\nonumber \\
&& -\frac{3}{2}(n_c+1)^2{\cal D}_{\alpha}^{(1)}  \left( \sum_{l=1}^{n_c}
\frac{W_{\alpha}(\tilde b_{l}+\tilde b_{l}^{\dagger})}{t_{l}(W_{\alpha}^2-\tilde \omega_{l}^2)} + \frac{\tilde b_{0}}{t_{0}W_{\alpha}} \right)
\left( \sum_{k=1}^{n_c}\frac{\tilde \omega_{k}(\tilde b_{k}^{\dagger} - \tilde b_{k})}{t_{k}(W_{\alpha}^2-\tilde \omega_{k}^2)}\right) \, , 
\eea
\bea
&&u_{1}^{\rm LH}=-\frac{2}{3}(n_c+1)^{\frac{3}{2}}\Bigl[ 2{\cal D}_{\alpha}^{(1)} \left( \sum_{k=1}^{n_c}
\frac{\tilde \omega_{k}B_{+}(\tilde b_{k}^{\dagger} -\tilde b_{k})}{t_{k}(W_{\alpha}^2-\tilde \omega_{k}^2)}\right) 
\nonumber \\
&&- \left(   \sum_{k=1}^{n_c} \frac{B_{-}(\tilde b_{k} + \tilde b_{k}^{\dagger})}{t_{k}}
 [2 {\cal D}_{\alpha}^{(2)}  + \frac{{\cal D}_{\alpha}^{(1)}W_{\alpha}}{W_{\alpha}^2-\tilde \omega_{k}^2}]
 + \frac{1}{t_{0}} B_{-}\tilde b_{0}(2{\cal D}_{\alpha}^{(2)} + \frac{{\cal D}_{\alpha}^{(1)}}{W_{\alpha}} )
 \right)\Bigr] \nonumber \\
&&  -\frac{4}{15}(n_c+1)^{\frac{5}{2}}{\cal D}_{\alpha}^{(1)}\left( {\cal D}_{\alpha}^{(2)} + \frac{{\cal D}_{\alpha}^{(1)} }{2W_{\alpha}} \right)
 \left( \sum_{k=1}^{n_c} \frac{W_{\alpha}B_{-}(\tilde b_{k} + \tilde b_{k}^{\dagger})}{t_{k}(W_{\alpha}^2-\tilde \omega_{k}^2)}   
 +\frac{B_{-}\tilde b_{0}}{t_{0}W_{\alpha}}  \right)
\eea
where 
\be
{\cal D}_{\alpha}^{(1)} = \sum_{k=1}^{n_c}\frac{2W_{\alpha}}{t_{k}^2(W_{\alpha}^2 - \tilde \omega_{k}^2)}  + \frac{1}{t_{0}^2W_{\alpha}} \, ,
\ee
\be
{\cal D}_{\alpha}^{(2)} = \sum_{k=1}^{n_c}\frac{\tilde \omega_{k}^2}{t_{k}^2(W_{\alpha}^2 - \tilde \omega_{k}^2)^2}\, .
\ee
For $\alpha\to \infty$ with $n_c$ fixed we have ${\cal D}_{\alpha}^{(1)}\sim 1/\alpha$ and ${\cal D}_{\alpha}^{(2)}  \sim 1/\alpha^4$.
This implies that the leading mixing interaction term in $u_{1}^{\rm LH}$ is of the order $1/\alpha^2$. We can remove it 
by a unitary transformation 
\be
h_{\alpha, n_c} \mapsto h_{\alpha, n_c}^{(2)} = e^{S_2}h_{\alpha, n_c}e^{-S_2}
\ee
 where $S_{2}$ is of the order $1/\alpha^3$. An inspection shows that this will introduce new interactions that start from the 
 order  $1/\alpha^5$. Hence all interactions to the order $1/\alpha^4$ are already contained in the formulas above.  
Expanding (\ref{u1L}) in powers of $1/\alpha$ we obtain the effective Hamiltonian for light modes 
\be\label{heffR}
h^{\rm eff}_{\rm IR} =  \tilde {\cal E}_{1}^{+}(\alpha, n_c) +  \sum_{k=1}^{n_c}\tilde \omega_{ k} \tilde b_{ k}^{\dagger}\tilde b_{ k}+ 
\tilde g^{\rm eff}(\alpha, n_c) \tilde T(0,0) + \tilde g^{\rm eff}_{3}(\alpha, n_c) \tilde {\cal O}_{3}(0,0) 
+ {\cal O}\left( \frac{1}{\alpha^5}\right)
\ee
where 
\be
\tilde T(0,0) = -\frac{\pi^2}{2} :\!
 \left( \sum_{l=1}^{n_c}\frac{\tilde b_{l}+\tilde b_{l}^{\dagger}}{t_{l}} + \frac{\tilde b_{0}}{t_{0}} \right)
\left( \sum_{k=1}^{n_c}\frac{\tilde \omega_{k}(\tilde b_{k}^{\dagger} - \tilde b_{k})}{t_{k}}\right)\!: \, , 
\ee
\be
\tilde {\cal O}_{3}(0,0)=-\frac{\pi^2}{2}
 :\!\left( \sum_{l=1}^{n_c}\frac{\tilde b_{l}+\tilde b_{l}^{\dagger}}{t_{l}} + \frac{\tilde b_{0}}{t_{0}} \right)
\left( \sum_{k=1}^{n_c}\frac{\tilde \omega_{k}^3(\tilde b_{k}^{\dagger} - \tilde b_{k})}{t_{k}}\right)\!: \, , 
\ee
and 
\be\label{geffIR}
\tilde g^{\rm eff}(\alpha, n_c)= -\frac{1}{2\pi^2\alpha^2} + \frac{3d^{(1)}_{n_c}}{8\pi^2\alpha^4} \, , 
\ee
\be
\tilde g^{\rm eff}_{3}(\alpha, n_c) = -\frac{1}{8\pi^2 \alpha^4 (n_c+1)} \, ,
\ee
\be
\tilde {\cal E}_{1}^{+}(\alpha, n_c)= \tilde {\cal E}_{1}(\alpha, n_c)-\frac{\pi^2}{2}\sum_{k=1}^{n_c}\frac{1}{t_{k}^2}
(g^{\rm eff}(\alpha, n_c)\tilde \omega_{k} + g^{\rm eff}_{3}(\alpha, n_c)\tilde \omega_{k}^3)\, .
\ee
Here 
\be
d^{(1)}_{n_c}=\sum_{k=1}^{n_c}\frac{1}{t_{k}^2} + \frac{1}{2t_{0}^2} \, .
\ee
In the large $n_c$ limit $t_{i} \to \pi$ so that we have 
\be
\tilde T(0,0) = T^{\, \rm tr}(0,0)  + {\cal O}(\frac{1}{n_c}) \, , \qquad 
\tilde {\cal O}_{3}(0,0) = {\cal O}_{3}^{\, \rm tr}(0,0)  + {\cal O}(\frac{1}{n_c})
\ee
where the operators on the right hand side are the stress energy tensor and $:\!\! \psi\partial_{\tau}^{3}\psi\!\!:$ 
operators regulated by mode truncation (see also (\ref{Otr_def})). For fixed $n_c$ we can thus consider $\tilde T$ 
and $\tilde O_{3}$ as the corresponding continuous operators regulated in some truncation scheme. 

The first term in (\ref{geffIR}) coincides with the leading irrelevant interaction found in \cite{Toth}. As for 
large $n_c$ we have 
\be 
d^{(1)}_{n_c} \to \frac{1}{\pi^2}(n_c + \frac{1}{2})
\ee
the second term in (\ref{geffIR}) is divergent in $n_c$ and can be interpreted as a counterterm stemming from 
the term in the OPE of stress-tensor $T$ with itself proportional to $T$:
\be
T(\tau,0)T(0, 0) \sim \frac{1}{4\tau^4} + \frac{2T(0,0)}{\tau^2} + \frac{\partial_{\tau}T(0,0)}{\tau} + \dots 
\ee
The corresponding divergence in the integrated correlator is not-integrable and being  regulated by mode truncation  gives a term proportional to $n_c$. 
Such a counterterm should be present in any truncation scheme.
In particular we find that  this counterterm removes the leading $n_c$-dependence in the spectral equation (\ref{g_spec2}) 
 with $g=\tilde g^{\rm eff}$.  Namely, substituting the leading asymptotic 
 \be
\tilde g^{\rm eff} =  -\frac{1}{2\pi^2\alpha^2} + \frac{3(n_c+1/2)}{8\pi^4\alpha^4}  
 \ee
into (\ref{g_spec2})  and expanding up to $1/\alpha^5$ we obtain 
\bea
&& \left[\sum_{k=0}^{n_c}\frac{2\omega^2}{n^2 - \omega^2} -1\right]^{-1}=-\frac{\tilde g^{\rm eff}(1+\tilde g^{\rm eff}(1+2n_c)/4)}
{(1+\frac{\tilde g^{\rm eff}}{2}(1+2n_c))^2} \nonumber \\
&&=-\tilde g^{\rm eff} + \frac{3}{4}(\tilde g^{\rm eff})^2(1+2n_c) + {\cal O}((\tilde g^{\rm eff})^3) \nonumber \\
&& = \frac{1}{2\pi^2\alpha^2} + {\cal O}\left( \frac{1}{\alpha^6}\right) \, .
\eea 
Noting that for $n_c \to \infty$ 
\be
\sum_{k=0}^{n_c}\frac{2\omega^2}{n^2 - \omega^2} -1 \to -\pi\omega\cot(\pi\omega)  
\ee
we reproduce the spectral equation (\ref{spec}) up to terms suppressed by $1/\alpha^6$ and by $1/n_c$. 
Note that the truncation scheme used in (\ref{g_Ham2}), (\ref{g_spec2}) is different from the mode truncation in the UV theory 
exported into IR which is used in (\ref{heffR}). 

At higher orders in $1/\alpha$ more derivative interactions ${\cal O}_{p}$ will appear. We observe  that the higher is the order 
of  derivative $p$ the higher is the leading suppressing power of $1/\alpha$ standing at   ${\cal O}_{p}$. 
 On general grounds, since only the vacuum energy diverges,  we expect that our effective infrared Hamiltonian contains 
 all the necessary counter terms to have a finite $n_c\to \infty$ limit of the excitation energies. 
 The leading interaction is given by the stress-energy tensor with a coupling 
independent of $n_c$. One may wonder whether the whole effective Hamiltonian should be considered 
as the leading interaction plus (infinitely many) counter terms which are needed to have a finite 
$n_c\to \infty$ limit. We note that in the present theory at higher order we generate more $n_c$-independent 
terms in the effective Hamiltonian. For example by expanding the last term in (\ref{u1L}) in powers of $\tilde \omega_{l}/W_{\alpha}$ 
we generate a finite term proportional to\footnote{Up to a total derivative that can be removed by a unitary transformation this operator 
is equivalent to ${\cal O}_{3}$.}
\be
\frac{1}{\alpha^6} :\!\! \partial_{\tau}^2 \psi \partial_{\tau}\psi \!\!: 
\ee
that suggests that this picture (leading order interaction plus counter terms) may be too simplistic\footnote{Strictly speaking we need to systematically perform  all effective Hamiltonian  calculations  to the order $1/\alpha^6$ to make sure this finite term does not get cancelled by other similar contributions.}. 

\section{Concluding remarks} \label{discussion_sec}
\setcounter{equation}{0}
Our main result regarding the RG interface is the explicit formula for the matrix elements (\ref{psi_hat1}). It would be interesting 
to calculate the normalisation factor ${\cal N}_{\alpha}$ or at least its asymptotic expansion at large $\alpha$. 
The vanishing of this prefactor is related to  the singularities in the OPE of   $\hat \psi_{0,\alpha}$ with itself. 

Another aspect of the RG interface which would be interesting to investigate is the transport of local operators. 
While the operator  $\hat \psi_{0,\alpha}$ establishes a mapping between states in the unperturbed and perturbed theories, 
 since at finite $\alpha$ the usual CFT state-operator correspondence does not work, this mapping does not automatically 
 establish a mapping of local operators.To establish the latter  we can follow the following procedure. 
Given an operator $\psi_{\rm UV}$
at the UV fixed point we can surround it by  $\hat \psi_{0,\alpha}$ and its conjugate taken some finite distance $\epsilon$ away 
from $\psi_{\rm UV}$ as 
depicted below.

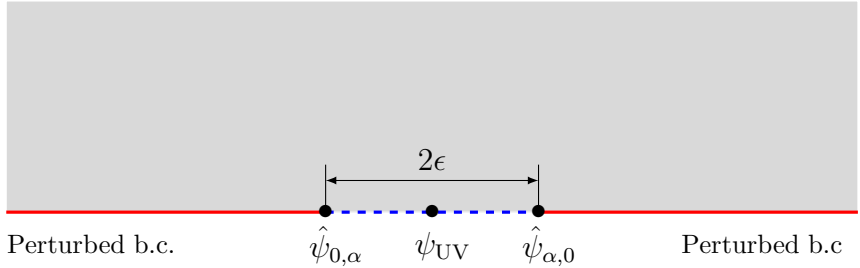
\begin{center}
\begin{tikzpicture}[>=latex,scale=1.4]
\filldraw[fill=gray!30!white,draw=white] (-4,0) rectangle (4,2);

\draw[blue,very thick,dashed](-1,0)--(1,0);
\draw[red,very thick] (-4,0)--(-1,0);
\draw[red,very thick] (1,0)--(4,0);
\draw (-1,0) node {$\bullet$} ;
\draw (1,0) node {$\bullet$} ;
\draw (0,0) node {$\bullet$};
\draw (1.1,-0.32) node {$\hat \psi_{\alpha, 0}$};
\draw (-0.9,-0.32) node {$\hat \psi_{0,\alpha}$};
\draw (0.1,-0.32) node {$\psi_{\rm UV}$};
\draw (-3.2,-0.3) node {{\small Perturbed b.c.}};
\draw (3.1, -0.3) node {{\small Perturbed b.c}} ;
\draw (-1,0) -- (-1, 0.45);
\draw (1,0)--(1,0.45);
\draw[<->] (-1, 0.3) -- (1,0.3);
\draw (0,0.5) node {$2\epsilon$};
\end{tikzpicture}
\end{center}

As we send $\epsilon $ to zero we should obtain, possibly after renormalisation, a local operator in 
the perturbed theory. Similarly we can start with a local operator in the perturbed theory and surround it 
by the interface operators distance $\epsilon$ away. In this case we would expect in the $\epsilon \to 0$ 
limit to obtain a local operator in the UV BCFT. It seems plausible that for finite $\alpha$ both procedures 
are the inverse of each other but it would be interesting to work out the details. We would further speculate   
that the limit $\alpha \to \infty$ does not commute with the limit $\epsilon \to 0$ as the OPE of the interface 
operator with itself and with other operators becomes singular at the infrared fixed point $\alpha=\infty$.  
Moreover we would expect that the two mappings: from UV operators to the IR operators and in the 
opposite direction, are no longer the inverses of each other. For the present model it should be possible to get 
precise answers to these questions that may enrich our intuition   about what happens to  local operators as we 
arrive at an infrared fixed point.

Regarding the second aspect of the paper -- the effective Hamiltonian near the IR fixed point, while it is instructing 
to see how it can be generated systematically term by term, it is desirable to understand better the role of regularisation 
scheme choice.  The mode truncation scheme we used is very special in two respects. Firstly, taking the coupling to 
infinity we obtain a finite spectrum approximating the low lying IR BCFT spectrum so that the truncation in the UV BCFT corresponds 
to some (local) truncation scheme in the IR BCFT. Secondly, for the perturbed truncated theory at finite value of 
the coupling the energies separate cleanly into low energy states and heavy energy states whose energies are 
all of the order $\alpha$. While everything works out fine for this regularisation scheme, the feeling of the author is that  its special features seem to 
obscure  some conceptual issues which remain to be uncovered. 

\vspace{1cm}



\appendix
\renewcommand{\theequation}{\Alph{section}.\arabic{equation}}
\setcounter{equation}{0}
\section{Explicit formulas for the  leg factors and matrices in the RG operator}
We start with the factorisation equation 
\be\label{factorisation2}
\Phi(\omega) \Phi(-\omega) = 1 + \frac{2\pi\alpha^2\tanh(\pi\omega)}{\omega} \, .
\ee
The function on the right hand side has simple poles at $\omega=i(k+1/2)$, $k\in {\mathbb Z}$ and simple zeroes 
at $\omega=\omega_{n}$, $n=\pm 1,\pm 2,\dots$ where $\omega_{n}$ are non-zero solutions to (\ref{spec}). 
In general (\ref{factorisation2}) defines $\Phi(\omega)$ up to a multiplication by $\exp(\omega A)$ where $A$ is a constant. 
We  choose 
\be\label{Phi}
\Phi(\omega) = \frac{\sqrt{1+2\pi^2\alpha^2}}{4\pi^{3/2}\alpha}\Gamma(1/2+i\omega)e^{i\gamma \omega}
\prod_{n=1}^{\infty}\left(1-\frac{\omega}{i\omega_{n}}\right)e^{-i\omega/n}
\ee
in such a way that for $|\omega| \to \infty$ 
\be
|\Phi(\omega)| \sim \frac{1}{\sqrt{|\omega|}} \, .
\ee
This gives 
 \be \label{dnn}
 d_{n} = 2\pi i {\rm Res} \tilde f_{+}(\omega) \Bigr|_{\omega= i(n+1/2)} = 
 \frac{\Gamma(n+1/2)}{\Gamma(n+1)}\frac{\sqrt{1+2\pi^2\alpha^2}}{2\alpha \pi^{3/2}}\prod_{k=1}^{\infty}
 \left( \frac{1-\frac{n+1/2}{\omega_{k}}}{1-\frac{n+1/2}{k}} \right) \, .
 \ee
We next take a logarithm of the infinite product and rewrite the corresponding series as a contour integral using a function
\be \label{W}
W(\omega) = \frac{2\pi^2\alpha^2}{2\pi^2\alpha^2 + 1}\left( 1 + \frac{\omega}{2\pi\alpha^2 \tanh(\pi\omega)}\right) 
\ee
that has simple poles at $\omega \in i{\mathbb Z}$ and simple zeroes at $\omega =\pm  i\omega_{k}$. The normalisation 
is chosen so that $W(0)=1$. For the logarithmic function rather than taking ${\rm Log}\left( 1 - \frac{i(n+1/2)}{\omega}\right) $, 
 that has a branch cut on the imaginary axis that passes through some zeroes and poles of $W(\omega)$, 
we choose ${\rm Log}(\omega-i(n+1/2))-{\rm Log}(\omega)$. Here and elsewhere  ${\rm Log}(z)$ stands for the principal branch function. 
With this in mind we write
\bea
&&P_{n}\equiv \prod_{k=1}^{\infty}
 \left( \frac{1-\frac{n+1/2}{\omega_{k}}}{1-\frac{n+1/2}{k}} \right)=\nonumber \\
 && \lim_{R\to \infty}\exp\Bigl[     \frac{1}{2\pi i} 
 \int\limits_{-R}^{R}dx\, ({\rm Log}(x-i(n+1/2))   -{\rm Log}(x))  d\ln W(x) \nonumber \\
&&  + \int\limits_{-R}^{0} dx\, d\ln W(x+i(n+1/2))  \Bigr]
\eea
where  the last term in the exponent comes from integrating along the 
branch cut of ${\rm Log}(x-i(n+1/2))$. Taking into account that $W(x)$ is even  the last expression simplifies to 
\be
P_{n} = \exp\left[ -\frac{1}{2\pi}\int\limits_{-\infty}^{\infty}\! dx\, {\rm arctan}\left(\frac{n+1/2}{x} \right) d\ln W(x)  
+ \ln\left(  \frac{2\pi^2\alpha^2}{2\pi^2\alpha^2+1} \right)  \right]
\ee
Integrating by parts and substituting $P_{n}$ into (\ref{dnn}) we finally obtain (\ref{dk}). 

Next we calculate the matrices $N_{nm}$ and $O_{pn}$ in (\ref{psi_hat1}). We have
\bea
&& O_{kn}=-i\langle 0|a_{k+1/2}\hat \psi_{0,\alpha} b^{\dagger}_{n}|0\rangle_{\alpha} =  \frac{2\alpha}{(k+1/2 -\omega_{n})f_{n}} 
-\frac{d_{k+1/2}}{f_{n}}  - \frac{2 \alpha d_{k+1/2}}{f_{n}}\nonumber \\
&& 
\Bigl(  -\frac{(2k+1)}{\omega_{n}-k-1/2}\sum_{r=0}^{\infty}\frac{d_{r}}{r+k+1}  
+ \left(\frac{\omega_{n}+k+1/2}{\omega_{n}-k-1/2}\right)\sum_{r=0}^{\infty}\frac{d_{r}}{r+1/2+\omega_{n}}\Bigr)
\eea
where we used the intertwining property of $\hat \psi_{0,\alpha}$. Using the recurrence relation (\ref{d_recurrence}) we recast the last expression 
as 
\be
O_{kn} = d_{k}\tilde d_{n} \left(\frac{k+1/2 + \omega_{n}}{k+1/2 - \omega_{n}} \right)
\ee
where 
\be\label{dtw}
\tilde d_{n} = \frac{1}{f_{n}}\left( 2\alpha \sum_{r=0}^{\infty}\frac{d_{r}}{r+1/2 + \omega_{n}} +1\right) \, .
\ee
Similarly using the intertwining property we obtain 
\be
\langle 0|a\hat \psi_{0,\alpha} b^{\dagger}_{n}|0\rangle_{\alpha} = \tilde d_{n} \, , 
\ee
\bea\label{N2}
&& N_{qp}=\langle 0|\hat \psi_{0,\alpha} b^{\dagger}_{p} b^{\dagger}_{q}|0\rangle_{\alpha} = -\tilde d_{q} \sum_{k=0}^{\infty} 
iB_{p,k}d_{k}\left(\frac{\omega_{q}+k+1/2}{\omega_{q}-k-1/2}\right) \nonumber \\
&& + \frac{1}{f_{p}}\left( \frac{1}{f_{q}}+ \sum_{r=0}^{\infty}i B_{q,r}d_{r}\right) \, . 
\eea
Furthermore, using 
\be
\sum_{r=0}^{\infty}\frac{id_{k}}{\omega_{q}-k-1/2} = \int\limits_{-\infty}^{\infty}\! d\omega\, \frac{\tilde f_{+}(\omega)}{\omega_{q}+i\omega} 
-2\pi \tilde f_{+}(i\omega_{q}) = \frac{1}{2\alpha}
\ee
we recast (\ref{N2}) as 
\be
N_{qp} = \tilde d_{q}\tilde d_{p} \frac{\omega_{p}-\omega_{q}}{\omega_{q}+\omega_{p}} \, . 
\ee

Formula (\ref{dtwiddle}) is obtained from (\ref{dtw}) by using 
\be
\sum_{r=0}^{\infty}\frac{d_{r}}{r+1/2 + \omega_{p}} = 2\pi \tilde f_{+}(-i\omega_{p}) 
\ee
and formula (\ref{Phi}) that give a product formula
\be
\tilde d_{p} = \frac{\sqrt{1+2\pi^2\alpha^2}\Gamma(\omega_{p}+1/2)}{f_{p}\sqrt{\pi}\Gamma(\omega_{p}+1)} 
\prod_{n=1}^{\infty}\left( \frac{1+\frac{\omega_{p}}{\omega_{n}}}{1+\frac{\omega_{p}}{n}}\right) \, .
\ee
Taking the logarithm of the infinite product factor and using a contour integration with  summation function (\ref{W}) we 
finally arrive at (\ref{dtwiddle}).

\setcounter{equation}{0}
\section{Some  identities used in  Schrieffer-Wolff calculations}

Here we collect some commutator formulas for the operators
\be
B_{+}=b_{\Omega} + b^{\dagger}_{\Omega} \, , \quad B_{-}=b_{\Omega}^{\dagger}- b_{\Omega} \, , \quad N=b_{\Omega}^{\dagger}b_{\Omega} \, .
\ee
These operators satisfy 
\be
B_{+}^2 = 1\, , \quad B_{-}^2 = -1 \, , \quad \{ B_{+}, B_{-}\} = 0 \, , 
\ee
\be
B_{-}B_{+} = 2N - 1\, , \qquad B_{+}B_{-}=-2N+ 1 \, , 
\ee
\be
[B_{+}, N]=-B_{-} \, , \qquad [B_{-}, N] = -B_{+} \, . 
\ee
We also calculate 
\be
[B_{+}(\tilde b_{n} +\tilde b_{n}^{\dagger}), B_{+}(\tilde b_{m} - \tilde b_{m}^{\dagger})]= -2(\tilde b_{n} +\tilde b_{n}^{\dagger})
(\tilde b_{m} - \tilde b_{m}^{\dagger}) \, , 
\ee
\be
[B_{+}(\tilde b_{n} +\tilde b_{n}^{\dagger}), B_{-}(\tilde b_{m} + \tilde b_{m}^{\dagger})]= 2(2N-1)\delta_{n,m}\, , 
\ee
\be
[B_{-}(\tilde b_{n} -\tilde b_{n}^{\dagger}), B_{-}(\tilde b_{m} + \tilde b_{m}^{\dagger})]= 2(\tilde b_{n} -\tilde b_{n}^{\dagger})
(\tilde b_{m} + \tilde b_{m}^{\dagger})\, , 
\ee
\be
[B_{-}(\tilde b_{n} -\tilde b_{n}^{\dagger}), B_{+}(\tilde b_{m} -\tilde  b_{m}^{\dagger})]=2(1-2N)\delta_{n,m}\, , 
\ee
\be
[B_{+}(\tilde b_{n} +\tilde b_{n}^{\dagger}),N]=-B_{-}(\tilde b_{n} +\tilde b_{n}^{\dagger})\, , \quad 
[B_{-}(\tilde b_{n} -\tilde b_{n}^{\dagger}), N]=-B_{+}(\tilde b_{n} -\tilde b_{n}^{\dagger}) \, , 
\ee
\be
[B_{+}(\tilde b_{k} +\tilde b_{k}^{\dagger}), (\tilde b_{n} +\tilde b_{n}^{\dagger})(\tilde b_{m} - \tilde b_{m}^{\dagger})]
= 2B_{+}\delta_{k,n}(\tilde b_{m} - \tilde b_{m}^{\dagger})\, , 
\ee
\be
[B_{-}(\tilde b_{k} -\tilde b_{k}^{\dagger}), (\tilde b_{n} +\tilde b_{n}^{\dagger})(\tilde b_{m} - \tilde b_{m}^{\dagger})]
= 2B_{-}\delta_{k,m}(\tilde b_{n} + \tilde b_{n}^{\dagger})\, .
\ee


\begin{thebibliography}{99}

\bibitem{GZ} S. Ghoshal and A. Zamolodchikov, {\it Boundary S-Matrix and Boundary State in Two-Dimensional Integrable Quantum Field Theory}, 
Int. J. Mod. Phys. {\bf A9} (1994) 3841-3886; Erratum-ibid. {\bf A9} (1994) 4353; arXiv:hep-th/9306002.
 

\bibitem{CZ} R. Chatterjee and A. Zamolodchikov, {\it Local Magnetization in Critical Ising Model with Boundary Magnetic Field},  
Mod. Phys. Lett. {\bf A9} (1994) 2227; arXiv:hep-th/9311165.

\bibitem{Chat} R. Chatterjee, {Exact Partition Function and Boundary State of Critical Ising Model with Boundary Magnetic Field}, 
Mod. Phys. Lett. {\bf A10} (1995) 973;  arXiv:hep-th/9412169.



\bibitem{me1}  A. Konechny, {\it Ising model with a boundary magnetic field - an example of a boundary flow},  JHEP {\bf 0412} (2004) 058; 
arXiv:hep-th/0410210.


\bibitem{Toth} G. Z. Toth, {\it A study of truncation effects in boundary flows of the Ising model on a strip}, 
 J. Stat. Mech. 0704 (2007) P04005; arXiv:hep-th/0612256.

\bibitem{Toth2} G. Z. Toth, {\it Investigations in Two-Dimensional Quantum Field Theory by the Bootstrap and TCSA Methods}, 
arXiv:0707.0015.

\bibitem{Chat2} R. Chatterjee, {\it Exact Partition Function and Boundary State of 2-D Massive Ising Field Theory with Boundary Magnetic Field}, 
Nucl. Phys. {\bf B468} (1996) 439; arXiv:hep-th/9509071.

\bibitem{Mir} O. Miroshnichenko, {\it Differential equation for local magnetization in the boundary Ising model}, 
Nucl. Phys. {\bf B811} ( 2009) 385; arXiv:0808.3808.
 
 



 \bibitem{BR} I. Brunner and D. Roggenkamp, {\it Defects and bulk perturbations of boundary 
  Landau-Ginzburg orbifolds}, JHEP {\bf 0804} (2008) 00; arXiv:0712.0188.
  
  \bibitem{FQ} S. Fredenhagen and T. Quella, {\it 	
Generalised permutation branes}, JHEP {\bf 0511} (2005) 004;  arXiv:hep-th/0509153.
  
\bibitem{Gaiotto} D. Gaiotto, {\it Domain walls for two-dimensional renormalization group flows}, JHEP 1212 (2012) 103;
arXiv:1201.0767.

\bibitem{me2} A. Konechny, {\it Renormalization group defects for boundary flows},  J. Phys. {\bf A46} (2013) 145401; arXiv:1211.3665.

\bibitem{TCSA_Ising} A. Konechny, {\it RG boundaries and interfaces in Ising field theory}, J. Phys. {\bf A50} (2017) no.14, 145403; 
arXiv:1610.07489. 




\bibitem{YZ} V. P. Yurov and Al. B. Zamolodchikov, {\it Correlation functions of integrable 2D models of the relativistic field theory: Ising model}, 
Int. J. of Mod. Phys. {\bf A6}, No. 19 (1991) 3419.  

\bibitem{FZ} P. Fonseca and A. Zamolodchikov, {\it 	
Ising field theory in a magnetic field: Analytic properties of the free energy}, Journal of Statistical Physics,
 Vol. 110, Issue 3 (2003) pp. 527-590; arXiv:hep-th/0112167.



\bibitem{Berezin} F. A. Berezin, {\it The method of second quantization}, Academic Press, 1966. 

\bibitem{Cardy} J. Cardy, {\it 	
Boundary Conditions, Fusion Rules and the Verlinde Formula }, Nucl. Phys. {\bf B324} (1989) 581-596. 

\bibitem{Cardy_var} J. Cardy, {\it Bulk Renormalization Group Flows and Boundary States in Conformal Field Theories}, 
SciPost Phys. 3,011 (2017);  arXiv:1706.01568.


\bibitem{BPPZ} R. E. Behrend, P.A. Pearce, V. B.  Petkova, and  J.-B. Zuber, {\it Boundary Conditions in Rational Conformal Field 
Theories}, Nucl. Phys. {\bf B579} (2000) 707-773; arXiv:hep-th/9908036.

\bibitem{Billo_etall} M. Billo, V. Gon\c calves, E. Lauria and M. Meineri, {\it Defects in conformal field theory}, 
JHEP {\bf 04} (2016) 091; ArXiv:1601.02883.


\bibitem{BulkBoundary} J. Armas and J. Tarrio, {\it On actions for (entangling) surfaces and DCFTs}, JHEP {\bf 1804} (2018) 100; arXiv:1709.06766.




 

\bibitem{SW} J. R. Schrieffer and P. A. Wolff, {\it Relation between the Anderson and Kondo Hamiltonians}, 
Phys. Rev. {\bf 149} (1966) p. 491.

\bibitem{Froechlich_etal} N. Datta, R. Fern\' andez, J. Fr\" ohlich and L. Rey-Bellet, 
{\it Low-temperature phase diagrams of quan- tum lattice systems. II. Convergent perturbation expansions and stability in systems with infinite degeneracy}, Helv. Phys. Acta {\bf 69} (1996)  pp.752?820.

\bibitem{SWreview} S. Bravyi, D. DiVincenzo and D. Loss, {\it Schrieffer-Wolff transformation for quantum many-body systems}, 
Ann. Phys. Vol. 326, No. 10 (2011) pp. 2793-2826; 	arXiv:1105.0675.



\bibitem{AlZ1} Al. B. Zamolodchikov, {\it Thermodynamic Bethe ansatz for RSOS scattering theories},  Nucl. Phys. {\bf B358} (1991) 497.

\bibitem{AlZ2} Al. B. Zamolodchikov, {\it From tricritical Ising to critical Ising by thermodynamic Bethe ansatz},  Nucl. Phys. {\bf B358} (1991) 524.

\bibitem{KM} T. Klassen and E. Melzer, {\it Spectral Flow between Conformal Field Theories in 1+1 dimensions}, Nucl. Phys. {\bf B370} (1992) 511.

\bibitem{Berkovich} A. Berkovich, {\it Conformal invariance, finite-size effects, and the exactcorrelators for the $\delta$-function Bose gas}, 
Nucl. Phys. {\bf B356} (1991) 655;

\bibitem{Feverati_etal} G. Feverati, E. Quattrini, and F. Ravanini, {\it Infrared Behaviour of Massless Integrable Flows entering the Minimal Models from $phi_{31}$ }, Phys. Lett. {\bf B374} (1996) 64; arXiv:hep-th/9512104.


\bibitem{spec_exact} E. E. Burniston and C. E. Siewert, {\it The use of Riemann problems in 
solving a class of transcendental equations}, Proc. Camb. Phil. Soc. {\bf 73} (1973) 111.


\bibitem{Lukyanov} S. Lukyanov, {\it Low energy effective Hamiltonian for the XXZ spin chain}, Nucl. Phys. {\bf B522} (1998) 533-549; 
 arXiv:cond-mat/9712314.
 
 \bibitem{LTerras} S. Lukyanov and V. Terras, {\it Long-distance asymptotics of spin-spin correlation functions for the XXZ spin chain}, 
 Nucl. Phys. {\bf B654} (2003) 323-356; 	arXiv:hep-th/0206093.
 
 \bibitem{Feverati_etal2} G. Feverati, K. Graham, P. A. Pearce, G. Zs. Toth, and G. Watts, {\it A Renormalisation group for TCSA}, 
 J. Stat. Mech. (2008) P03011; 	arXiv:hep-th/0612203.
 
\bibitem{Gerard2}   G. Watts, {\it On the renormalisation group for the boundary Truncated Conformal Space Approach}, 
Nucl. Phys. {\bf B859} (2012) 177-206; arXiv:1104.0225.

\bibitem{Zam_TT} F.A. Smirnov, A.B. Zamolodchikov, {\it On space of integrable quantum field theories}, 
Nucl. Phys.  {\bf B915} (2017) pp. 363-383;  arXiv:1608.05499.



\bibitem{Tateo_etal} A. Cavagli\`a, S. Negro, I. M. Sz\' ecs\' enyi, and R. Tateo, {\it $T\bar T$-deformed 2D Quantum Field Theories}, 
JHEP {\bf 10} (2016) 112;  arXiv:1608.05534.

\bibitem{KM2} A. Konechny and D. McAteer, {On asymptotic behaviour in TCSA}, in preparation. 

\bibitem{AK_inprogress} A. Konechny, {\it Boundary renormalisation group  interfaces}, work in progress. 

\end{thebibliography}
\end{document}